\documentclass[sii]{ipart}
\usepackage{hyperref}
\usepackage{indentfirst}
\usepackage{enumitem}
\usepackage{multirow}
\usepackage{array}
\usepackage{nccmath}
\usepackage{subcaption}
\usepackage{caption}
\usepackage{graphicx}
\usepackage{booktabs}
\usepackage{amsthm}

\startlocaldefs
\theoremstyle{plain}

\theoremstyle{definition}
\newtheorem{definition}{Definition}[section]
\DeclareMathOperator{\cl}{cl}
\DeclareMathOperator{\CL}{CL}
\endlocaldefs

\newcolumntype{C}{>{\centering\arraybackslash}p{1.8cm}}

\DeclareCaptionLabelSeparator{quadsep}{\quad}
\pubyear{2026}
\volume{0}
\issue{0}
\firstpage{1}
\lastpage{0}
\arxiv{}

\begin{document}

\begin{frontmatter}

\title{Statistical Depth for Interval Data: A Closeness-Based Approach and Its Applications}

\begin{aug}
\author{\inits{X.~Z.}\fnms{Xiaozheng} \snm{CHEN} }
\address{Renmin University of China, Beijing, China\\
%\printead{wenlin.dai@ruc.edu.cn}
}
and
\author{\inits{W.~L.}\fnms{Wenlin} \snm{DAI$^*$}\ead[label=e1]{wenlin.dai@ruc.edu.cn}}
\address{Renmin University of China, Beijing, China\\\printead{e1}
}
\thankstext{t1}{Corresponding author.}
\end{aug}

\received{}

\begin{abstract}
With the rapid development of technology, interval data have been increasingly widely used in various fields, as they can effectively carry information on measurement uncertainty and original variability. Reasonable ranking and outlier screening of interval data are crucial for exploring their value, and statistical depth theory is an important tool to achieve this goal. Existing statistical depth methods are mostly defined for ordinary point-valued data, which are difficult to adapt to the characteristics of interval data and cannot meet the needs of their ranking and outlier detection. Therefore, this paper proposes a statistical depth suitable for interval data and conducts relevant application verification.
\end{abstract}


\end{frontmatter}

\section{INTRODUCTION}
With the rapid growth in dataset size and complexity, statistical analysis has evolved into a cornerstone of modern research. Traditional statistical methods relying on point-valued data, however, face significant limitations when dealing with inherently uncertain measurements—from sensor drifts in environmental monitoring to confidence intervals in clinical trials. Interval data, where observations are represented as bounded intervals (e.g., daily temperature $[20^\circ\text{C}, 25^\circ\text{C}]$), provide a natural framework to encode such uncertainties while preserving the original variability structure. As demonstrated in geospatial analysis and financial risk modeling \cite{Billard2003, Brito2006}, interval-based approaches are more capable of capturing intrinsic data variability in contexts requiring explicit uncertainty quantification, compared with traditional point-valued data methods.
The statistical analysis of interval data has consequently emerged as a critical research frontier. Symbolic data analysis (SDA), a paradigm extending classical tools to complex data structures, offers systematic techniques for interval datasets. Unlike conventional methods that summarize data through scalar statistics, SDA operates on multidimensional symbolic objects such as hypercubes \cite{BOCK2000} and employs interval matrices for algebraic operations \cite{BILLARD2008}, enabling a more faithful representation of multidimensional relationships. This shift necessitates novel inferential frameworks addressing both global patterns and local boundary behaviors inherent to interval data.

A particularly promising direction lies in extending statistical depth — a multivariate centrality measure pioneered by \cite{Tukey1975} with his halfspace depth — to interval data contexts. Though depth functions have revolutionized anomaly detection and nonparametric inference for point data \cite{Zuo2000}, their adaptation to interval systems requires solving unique challenges. Beyond simply ranking intervals by midpoint positions, meaningful depth must account for interaction effects between interval widths (uncertainty ranges) and their spatial configurations—a problem currently lacking systematic solutions.

Efforts to advance interval data analysis have yielded specialized methodologies across diverse domains. In spatial and environmental analytics, \cite{Freitas2022} dedicated to the geospatial analysis of interval data and extended it to Moran’s spatial autocorrelation index, which proves effective in quantifying uncertainty propagation within regional-scale climate ensembles. More recently, \cite{Workman2023} proposed geospatial analysis methods for interval data and validated their performance through a case study on geospatial variability, with potential applications in soil contaminant prediction by preserving interval boundaries and enriching interval-based environmental modeling techniques. In computational econometrics, \cite{SUN2022772} developed a model averaging framework for interval-valued time series, which leverages interval time-series structures and incorporates interval-valued exogenous variables to enhance crude oil price volatility forecasts via scientific weighting mechanisms. In data quality assurance, \cite{Silva2018} focused on outlier detection in interval data and introduced a parametric system for high-dimensional financial transaction intervals, grounded in multivariate Mahalanobis distance distributions optimized to account for variable correlations.

While these discipline-specific contributions to interval data analysis highlight its analytical potential, they collectively reveal two systemic limitations: (1) methodological frameworks remain closely tied to their originating domains, hindering cross-disciplinary knowledge synthesis; (2) foundational statistical paradigms, particularly depth function theory, lack a consistent interval-based axiomatization that would enable robust multidimensional inference.

This paper proposes to bridge these gaps by developing a statistical depth framework specifically tailored for interval data. First, we establish the theoretical foundations of depth functions for interval observations; second, we validate their practical efficacy in anomaly detection and time-series modeling. This study seeks to advance the theoretical understanding of interval data while developing practical tools to capture its unique characteristics.

The remainder of this paper is organized as follows. In Section 2, we review the theoretical background of statistical depth and symbolic data analysis. Section 3 introduces our proposed framework for interval data analysis, supported by illustrative examples. Section 4 discusses applications in anomaly detection and time-series modeling, while Section 5 concludes with insights and future research directions.

\section{BACKGROUND AND NOTATION}

\subsection{Interval data}
In this section, we formalize the fundamental concepts of interval-valued data and contextualize its analytical relevance relative to classical point-valued data. In classical statistical analysis, a dataset is typically represented as an $n \times p$ matrix, where each row corresponds to one of $n$ samples and each column denotes a single-valued variable with a scalar observation. Symbolic data analysis (SDA) extends this framework by allowing multi-valued (and weighted) entries, with interval-valued data being a core subclass that naturally encodes measurement uncertainty and variability.

An interval-valued variable takes values as closed real intervals rather than scalars. For instance, age can be categorized as interval-valued (e.g., $[0,10)$, $[10,20)$, $[20,30)$, $\dots$) and blood pressure can be represented as a range (e.g., $\xi_{ij} = [78,120]$). Following \cite{Brito2006}, an interval-valued random variable is defined as a mapping $Y: \Omega \to \mathcal{I}(\mathbb{R})$, where $\Omega = \{ \omega_1, \dots, \omega_n \}$ denotes the sample space and $\mathcal{I}(\mathbb{R})$ is the set of all closed intervals on the real line $\mathbb{R}$. For each sample $\omega_i \in \Omega$, $Y(\omega_i) = [l_i, u_i]$ with $l_i \leq u_i$, where $l_i$ and $u_i$ are the lower and upper bounds of the interval, respectively.

Formally, an $n \times p$ interval-valued dataset is represented by the matrix $\mathbf{I}$, where rows correspond to $n$ samples and columns to $p$ interval-valued variables:
\begin{equation*}
    \mathbf{I} = 
    \begin{pmatrix}
        [ l_{11}, u_{11} ] & \cdots & [ l_{1p}, u_{1p} ] \\
        \vdots & \ddots & \vdots \\
        [ l_{n1}, u_{n1} ] & \cdots & [ l_{np}, u_{np} ]
    \end{pmatrix}
    \label{eq:interval_matrix}
\end{equation*}
Here, $[ l_{ij}, u_{ij} ] \subseteq \mathbb{R}$ denotes the interval observation of the $j$-th interval-valued variable $Y_j$ ($j=1,\dots,p$) for the $i$-th sample $\omega_i$ ($i=1,\dots,n$), with the constraint $l_{ij} \leq u_{ij}$ for all $i,j$.

The $i$-th sample $\omega_i$ is equivalently expressed as a row vector in the $p$-dimensional interval space $\mathbb{IR}^p$:
\begin{equation*}
    \mathbf{I}_i = \left( \mathopen{[} l_{i1}, u_{i1} \mathclose{]}, \dots, \mathopen{[} l_{ip}, u_{ip} \mathclose{]} \right) \in \mathbb{I}\mathbb{R}^p
    \label{eq:interval_sample_vector}
\end{equation*},
where $\mathbb{IR}^p = \underbrace{\mathcal{I}(\mathbb{R}) \times \dots \times \mathcal{I}(\mathbb{R})}_{p \text{ times}}$ denotes the space of $p$-dimensional interval vectors.

\subsection{Statistical depth}
Statistical depth quantifies the centrality of an observation within a probability distribution, serving as a foundational tool for robust central tendency estimation and anomaly detection in multivariate data. First introduced by \cite{Tukey1975}, the concept of statistical depth originated with the \textbf{half-space depth} (HSD), which extends univariate rank and order statistics to the multivariate setting via depth-induced contours. Key advancements in depth theory include:

\begin{itemize}
    \item \cite{Liu1990} formalized the \textbf{simplicial depth} using affine geometric principles, and for the first time established four axiomatic properties that a valid statistical depth function should satisfy;
    \item \cite{Zuo2000} unified disparate depth functions into a consistent theoretical framework and rigorously evaluated their compliance with core mathematical axioms.
\end{itemize}

\begin{definition}[{\citeauthor{Zuo2000} \citeyear{Zuo2000}; \citeauthor{Liu1990} \citeyear{Liu1990}}]
\label{def:depth}
A bounded, nonnegative functional 
\[
D(\cdot, \cdot): \mathbb{R}^p \times \mathcal{P} \rightarrow \mathbb{R}
\] 
is termed a \textbf{statistical depth function} if it satisfies the following four axiomatic properties for all probability distributions $P \in \mathcal{P}$ (the set of all $p$-dimensional probability distributions) and all $\mathbf{x} \in \mathbb{R}^p$:
\begin{enumerate}
    \item \textbf{Affine invariance}: For any nonsingular matrix $\mathbf{A} \in \mathbb{R}^{p \times p}$ and translation vector $\mathbf{b} \in \mathbb{R}^p$,
        \[
        D\big(\mathbf{A}\mathbf{x} + \mathbf{b}, P_{\mathbf{A}\mathbf{X} + \mathbf{b}}\big) = D\big(\mathbf{x}, P_{\mathbf{X}}\big),
        \]
        where $\mathbf{X}$ denotes a $p$-dimensional random vector following distribution $P_{\mathbf{X}}$.
    
    \item \textbf{Maximality at the center}: If $P$ has a unique symmetry center $\boldsymbol{\theta} \in \mathbb{R}^p$, then
        \[
        D(\boldsymbol{\theta}, P) = \sup_{\mathbf{x} \in \mathbb{R}^p} D(\mathbf{x}, P).
        \]
    
    \item \textbf{Monotonicity relative to the deepest point}: Let $\boldsymbol{\theta}$ be the deepest point of $P$ (i.e., $D(\boldsymbol{\theta}, P) = \sup_{\mathbf{x} \in \mathbb{R}^p} D(\mathbf{x}, P)$). For all $\alpha \in [0,1]$,
        \[
        D(\mathbf{x}, P) \leq D\big(\boldsymbol{\theta} + \alpha(\mathbf{x} - \boldsymbol{\theta}), P\big).
        \]
    
    \item \textbf{Vanishing at infinity}:
        \[
        \lim_{\|\mathbf{x}\| \to \infty} D(\mathbf{x}, P) = 0.
        \]
\end{enumerate}
\end{definition}

Among widely used depth functions, the half-space depth and its generalization (projection depth) exhibit exceptional robustness for multivariate anomaly detection, as they are less sensitive to outliers compared to traditional moment-based methods.

\section{FORMAL DEFINITION OF INTERVAL DEPTH}

\subsection{A Closeness Measure Between Point and Interval}
In the field of data analysis, point-to-point relationships have been well characterized
by mature distance metrics such as the Euclidean distance.
For interval-to-interval relationships, set-theoretic operations including intersection and union
are commonly used to characterize their positional relations.
Nevertheless, a fundamental and still open question remains:
how to formally quantify the closeness between a single point and an interval.

While the containment relationship between a point and an interval can be easily determined, there is no dedicated measure to evaluate their closeness when the point lies outside the interval. To address this gap, we propose a closeness measure that quantifies the positional affinity between a point and an interval, regardless of whether the point is contained within the interval or not.

\begin{definition}[Point-Interval Closeness]
\label{def:point_interval_closeness}
Given a scalar point $x \in \mathbb{R}$ and a non-degenerate closed interval $I = [a, b]$ (with $a < b$), the closeness of $x$ to $I$ is defined as:
\begin{equation}
\label{eq:point_interval_closeness}
\cl(I, x) = \frac{b - a}{|x - b| + |x - a|}
\end{equation},
where $a$ and $b$ are the lower and upper bounds of $I$, respectively.
\end{definition}

To visually illustrate the intuitive behavior of $\cl(I, x)$, we take the symmetric interval $I = [-1, 1]$ as a canonical example and plot the variation of $\cl(I, x)$ with respect to the position of $x$. This example eliminates the distraction of arbitrary interval endpoints and highlights the core trend of the measure.

\begin{figure}[ht]
\centering
\includegraphics[width=5cm]{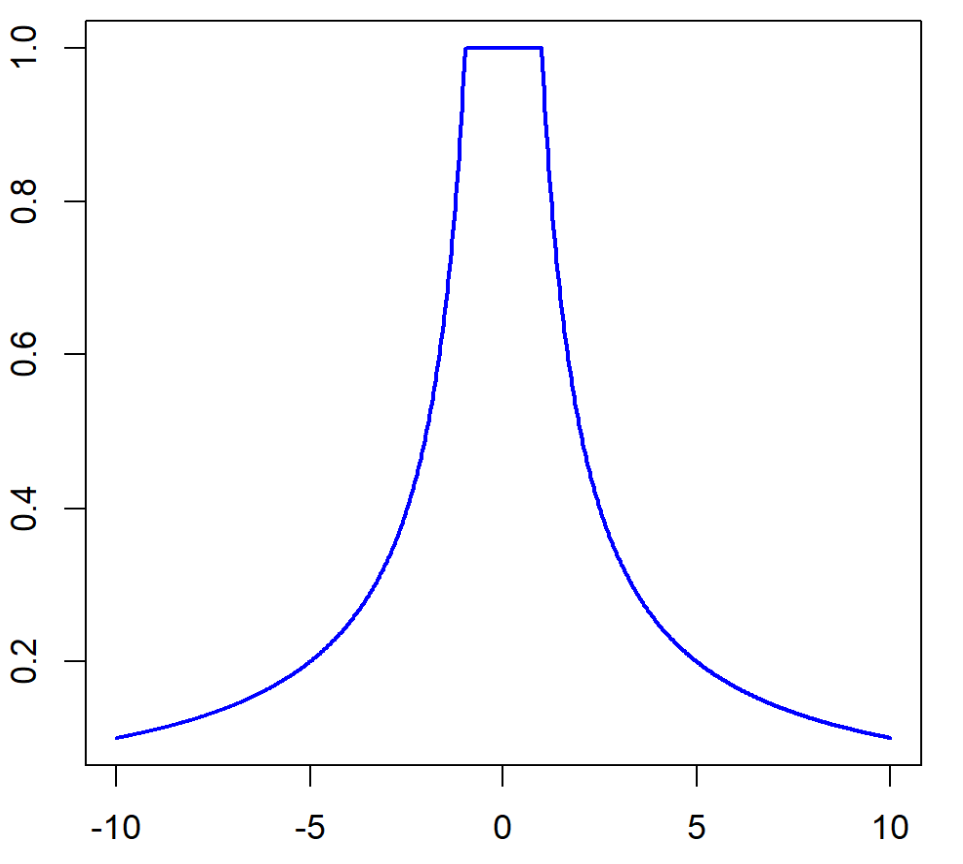}
\caption{Variation trend of point-to-interval closeness $\cl(I, x)$ for the symmetric interval $I = [-1, 1]$.}
\label{fig:point_interval_cl}
\end{figure}

As shown in Figure \ref{fig:point_interval_cl}, $\cl(I, x)$ exhibits the key characteristics defined in the properties below: it maintains a constant maximum value of 1 for all points within $I$, and decays smoothly to 0 as $x$ moves to positive or negative infinity. This visual pattern directly validates the bounded interpretability of the measure and provides a geometric intuition for its design.

The design of $\cl(I, x)$ is justified by three intuitive properties that align with the requirements of interval data analysis:

\begin{enumerate}
\item \textit{Bounded Interpretability.}
$\cl(I, x) \in (0, 1]$. The maximum value $\cl(I, x) = 1$ is achieved if and only if $x \in [a, b]$ (the point is fully contained in the interval), which directly quantifies containment. For $x \notin [a, b]$, the value decreases as the point moves away from the interval, reflecting diminishing closeness.
Formally,
\[
\cl(I, x) \in (0, 1] \quad \text{and} \quad \cl(I, x) = 1 \iff x \in [a, b].
\]

\item \textit{Length Sensitivity.}
The numerator $b-a$ (interval length) ensures that the measure accounts for the "size" of the interval—unlike point-point distance metrics that ignore interval characteristics. A point near a longer interval will have a higher closeness value than the same point near a shorter interval at the same spatial distance, which is consistent with practical intuition.

\item \textit{Scale and Translation Invariance.}
For any scalar $\alpha > 0$ and translation $\beta \in \mathbb{R}$, the measure remains invariant under uniform scaling and global translation. This guarantees consistent results across different measurement scales and coordinate systems, a necessary property for statistical depth axiomatization.
Formally, for $\alpha > 0$ and $\beta \in \mathbb{R}$, let $\alpha I + \beta = [\alpha a + \beta, \alpha b + \beta]$, then
\[
\cl(\alpha I + \beta, \alpha x + \beta) = \cl(I, x).
\]
\end{enumerate}

\subsection{One-Sided Interval Closeness}
\label{subsec:one_sided_closeness}

Following the definition of the point-to-interval closeness measure (Definition \ref{def:point_interval_closeness}), we extend this concept to characterize the closeness between two intervals. Specifically, to quantify the closeness between $I_1 = [a, b]$ and $I_2 = [c, d]$, we compute the normalized integral average of the point-to-interval closeness $\cl(I_1, x)$ (for each point $x\in I_2$) over the domain of $I_2$, effectively aggregating the point-wise proximity metric into an interval-level measure.

Accordingly, the closeness from interval $I_2$ to $I_1$, denoted $\cl(I_1, I_2)$, is formulated as:
\[
\cl(I_1, I_2) = \frac{1}{d - c} \int_{c}^{d} \frac{b - a}{|x - b| + |x - a|} \, dx
\]

We first consider the disjoint case where \(a < b < c < d\) (i.e., $I_1$ lies entirely to the left of $I_2$). For this scenario, the absolute-value term in the integrand can be simplified, leading to:
\[
\cl(I_1, I_2) = \frac{b - a}{2(d-c)} \int_{c}^{d} \frac{1}{x - \frac{a+b}{2}} dx
\],

Let $m = \frac{a+b}{2}$ denote the midpoint of $I_1$. Evaluating the definite integral yields:
\[
\cl(I_1, I_2) = \frac{b - a}{2(d-c)} \left[ \ln \left| x - m \right| \right]_{c}^{d} = \frac{b - a}{2(d-c)} \ln\left( \frac{d - m}{c - m} \right)
\]
(Note: The absolute-value sign is omitted here since $d>c>b>m$, ensuring $d-m>0$ and $c-m>0$.)

For overlapping intervals, we adopt a piecewise integration strategy: the integral domain is partitioned according to the intersection of $I_1$ and $I_2$, and a weighted average of the integral results over each subdomain is computed. Below, we present a comprehensive definition of $\cl(I_1, I_2)$ that covers all possible positional relationships between $I_1$ and $I_2$.

\begin{definition}
Given two non-degenerate closed intervals $I_1=[a,b]$ and $I_2=[c,d]$, the one-sided closeness from $I_2$ to $I_1$, denoted $\cl(I_1,I_2)$, is defined as follows:

%\resizebox{\linewidth}{!}{%
%\begin{equation}
%\cl(I_1,I_2) =
%\begin{cases} 
%\dfrac{1}{2}\cdot\dfrac{b-a}{d-c}\left| \ln\dfrac{c-m}{d-m} \right|,
%& \text{if } a<b\le c<d, \\
%& \text{or } c<d\le a<b, \\[1em]
%
%\dfrac{1}{d-c}\left[ b-c + \dfrac{1}{2}(b-a)\left| \ln\dfrac{b-m}{d-m} \right| \right],
%& \text{if } a<c<b<d, \\[1em]
%
%\dfrac{1}{d-c}\left[ d-a + \dfrac{1}{2}(b-a)\left| \ln\dfrac{c-m}{a-m} \right| \right],
%& \text{if } c<a<d<b, \\[1em]
%
%1,
%& \text{if } a<c<d<b, \\[1em]
%
%\dfrac{1}{d-c}\bigg[ b-a + \dfrac{b-a}{2}\left| \ln\dfrac{b-m}{d-m} \right| + \dfrac{b-a}{2}\left| \ln\dfrac{c-m}{a-m} \right| \bigg],
%& \text{if } c<a<b<d,
%\end{cases}
%\end{equation}
%}

\begin{equation}
	\resizebox{\linewidth}{!}{$
		\cl(I_1,I_2) =
		\begin{cases} 
			\dfrac{1}{2}\cdot\dfrac{b-a}{d-c}\left| \ln\dfrac{c-m}{d-m} \right|,
			& \text{if } a<b\le c<d, \\
			& \text{or } c<d\le a<b, \\[1em]
			
			\dfrac{1}{d-c}\left[ b-c + \dfrac{1}{2}(b-a)\left| \ln\dfrac{b-m}{d-m} \right| \right],
			& \text{if } a<c<b<d, \\[1em]
			
			\dfrac{1}{d-c}\left[ d-a + \dfrac{1}{2}(b-a)\left| \ln\dfrac{c-m}{a-m} \right| \right],
			& \text{if } c<a<d<b, \\[1em]
			
			1,
			& \text{if } a<c<d<b, \\[1em]
			
			\dfrac{1}{d-c}\bigg[ b-a + \dfrac{b-a}{2}\left| \ln\dfrac{b-m}{d-m} \right| + \dfrac{b-a}{2}\left| \ln\dfrac{c-m}{a-m} \right| \bigg],
			& \text{if } c<a<b<d,
		\end{cases}
		$}
\end{equation}

where $m=\frac{a+b}{2}$ is the midpoint of interval $I_1$.
\end{definition}

The one-sided interval closeness measure $\cl(I_1, I_2)$ satisfies the following intuitive properties, which validate its rationality for interval data analysis:

\begin{enumerate}
\item \textit{Boundedness.}
$\cl(I_1, I_2) \in (0, 1]$. The maximum value 1 is achieved if and only if $I_2$ is fully contained in $I_1$ (i.e., $a < c < d < b$), which directly reflects complete closeness between the two intervals. For non-overlapping intervals, $\cl(I_1, I_2)$ decreases as the distance between $I_1$ and $I_2$ increases, consistent with practical intuition.

\item \textit{One-Sided Asymmetry.}
In general, $\cl(I_1, I_2) \neq \cl(I_2, I_1)$. This asymmetry is intentional and aligns with the definition of one-sided closeness: $\cl(I_1, I_2)$ quantifies how close $I_2$ is to $I_1$, while $\cl(I_2, I_1)$ quantifies the reverse relationship.

\item \textit{Containment Invariance.}
If $I_2 \subseteq I_1$ (i.e., $a < c < d < b$), then $\cl(I_1, I_2) = 1$ for any valid interval endpoints. This property ensures that the measure correctly identifies full containment as the maximum closeness.

\item \textit{Continuity.}
$\cl(I_1, I_2)$ is continuous with respect to the endpoints $a, b, c, d$ (excluding degenerate intervals where $a=b$ or $c=d$). This continuity guarantees that small perturbations to the interval boundaries do not cause abrupt changes in the closeness value, a critical property for robust statistical analysis.
\end{enumerate}

To further illustrate the behavior of $\cl(I_1, I_2)$ across different structural characteristics of $I_2$, we fix $I_1 = [-1, 1]$ (i.e., $a=-1, b=1, m_1=0$) and analyze how $\cl(I_1, I_2)$ varies with two core parameters of $I_2$: its midpoint $m_2$ and its length $l_2$ . Figure \ref{fig:cl_parameter_variation} presents two subfigures that characterize this variation:

\begin{figure}[ht]
\centering
\begin{subfigure}{0.48\linewidth}
    \centering
    \includegraphics[width=\linewidth]{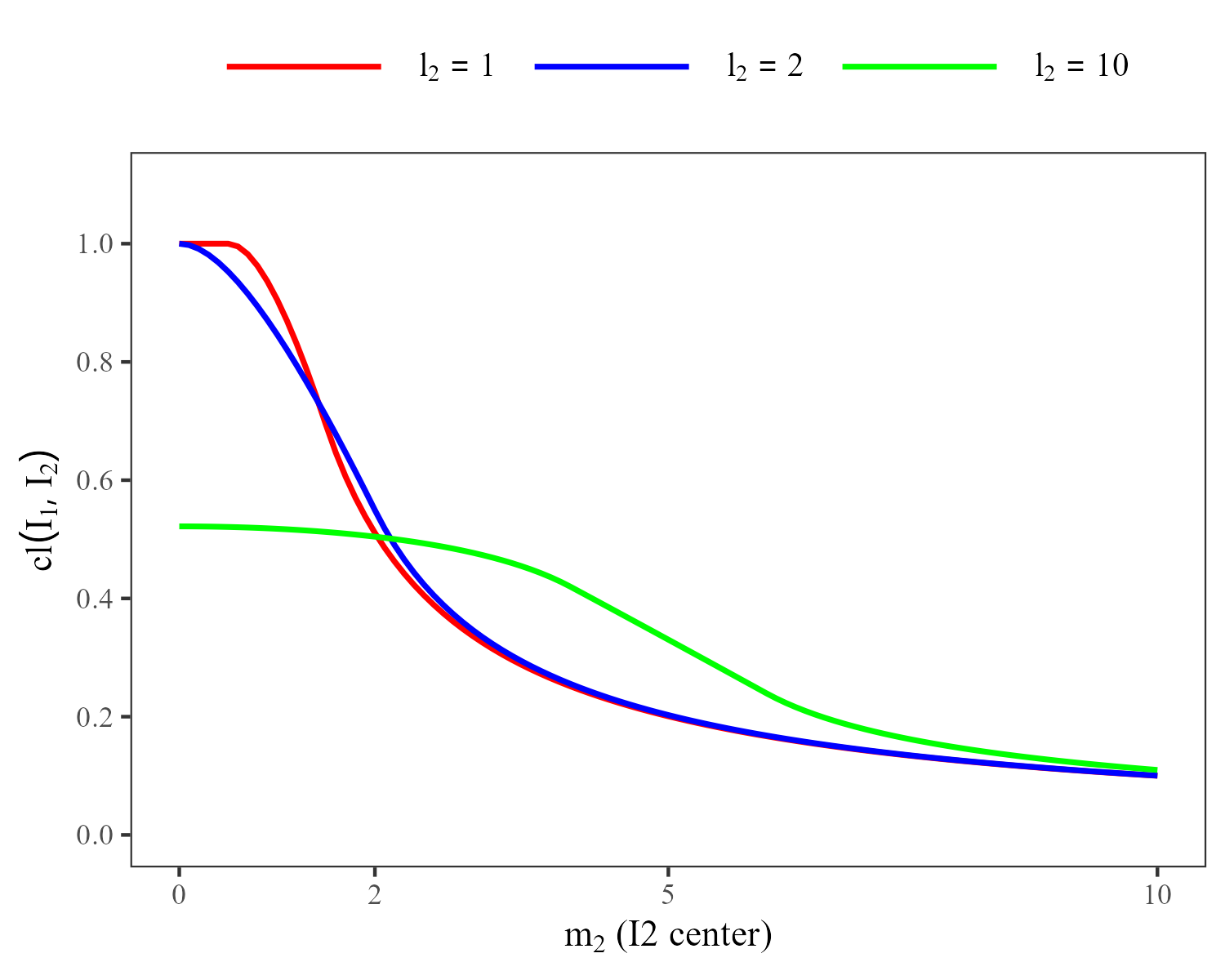}
    \caption{Fixed $l_2$, varying $m_2$}
    \label{subfig:cl_m2_variation}
\end{subfigure}
\hfill
\begin{subfigure}{0.48\linewidth}
    \centering
    \includegraphics[width=\linewidth]{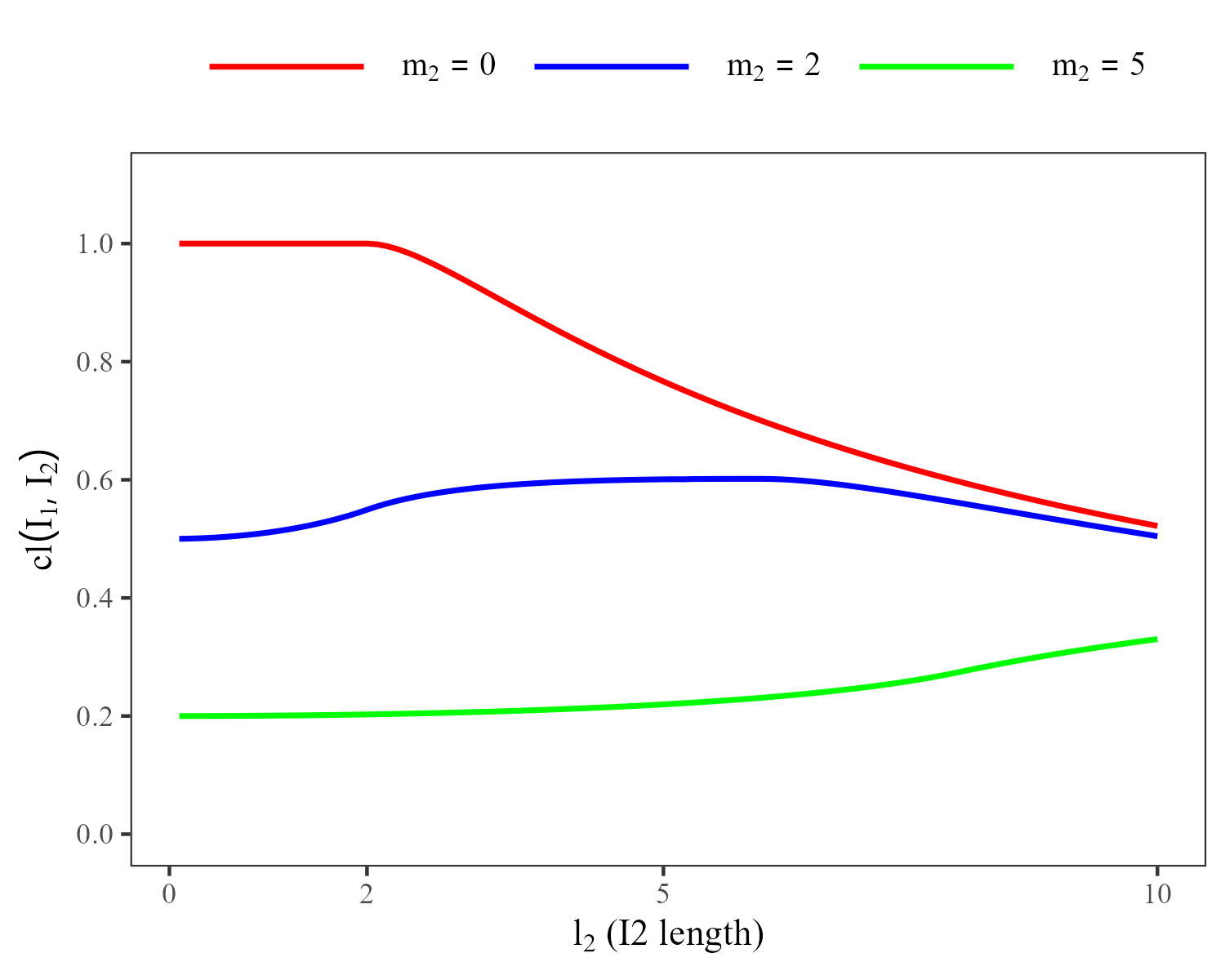}
    \caption{Fixed $m_2$, varying $l_2$}
    \label{subfig:cl_l2_variation}
\end{subfigure}
\caption{Variation of $\cl(I_1, I_2)$ with $m_2$ (midpoint of $I_2$) and $l_2$ (length of $I_2$), where $I_1 = [-1,1]$.}\label{fig:cl_parameter_variation}
\end{figure}

Subfigure \ref{subfig:cl_m2_variation} examines the impact of positional shift ($m_2$) on $\cl(I_1, I_2)$ for three representative lengths of $I_2$ ($l_2=1, 2, 10$). The results confirm \textit{Boundedness} and \textit{Continuity}: $\cl(I_1, I_2)$ remains at 1 when $I_2$ is fully contained in $I_1$ (e.g., $m_2 \in [-0.5, 0.5]$ for $l_2=1$), and decreases continuously as $m_2$ moves away from $I_1$. Longer intervals ($l_2=10$) exhibit slower decay in closeness because their larger spatial extent maintains partial overlap with $I_1$ even at greater positional shifts.

The figure \ref{subfig:cl_l2_variation} further analyzes the effect of the interval length ($l_2$) on $\cl(I_1, I_2)$ for three fixed midpoints ($m_2=0, 2, 5$). For $m_2=0$ (aligned with $I_1$), $\cl(I_1, I_2)$ remains 1 for all $l_2$ that keep $I_2 \subseteq I_1$, directly validating \textit{Containment Invariance}. For spatially separated intervals ($m_2=2, 5$), increasing $l_2$ expands $I_2$ to overlap with $I_1$, gradually restoring $\cl(I_1, I_2)$ toward 1—demonstrating that length acts as a "compensation factor" for positional separation. Together, these results confirm that $\cl(I_1, I_2)$ behaves consistently with the theoretical properties defined above, effectively quantifying one-sided interval closeness across all positional and structural configurations of $I_2$.

\subsection{Symmetric Normalized Closeness}
\label{subsec:sym-closeness}

The one-sided interval closeness captures asymmetric relationships between intervals. Specifically, the measure \(\mathrm{cl}(I_1, I_2)\) differs from \(\mathrm{cl}(I_2, I_1)\) due to the combined effects of interval lengths and positional offsets. This asymmetry introduces inherent biases when the two intervals \(I_1 = [a, b]\) and \(I_2 = [c, d]\) (consistent with the notation in the one-sided closeness section) have significantly different lengths (\(b - a \neq d - c\)), limiting its applicability in symmetric analysis tasks (e.g., interval similarity assessment, clustering).

To eliminate length-induced biases and adapt the measure to practical symmetric scenarios, we construct a symmetric normalized closeness measure via length-weighted averaging of the two one-sided closeness values:
\begin{equation}\label{eq:CL-def}
\begin{aligned}
    \CL(I_1, I_2) = \frac{ \cl(I_1, I_2) \cdot |I_2| + \cl(I_2, I_1) \cdot |I_1| }{ |I_1| + |I_2| },
\end{aligned}
\end{equation}
where $|I_1| = b - a$ and $|I_2| = d - c$ denote the lengths of $I_1=[a,b]$ and $I_2=[c,d]$, respectively.

The denominator \((b - a) + (d - c)\) represents the combined length of \(I_1\) and \(I_2\), while the numerator linearly weights each one-sided closeness by the length of the counterpart interval—this weighting scheme effectively mitigates biases caused by unequal interval lengths.

Three key properties of \(\mathrm{CL}\) are derived as follows:
\begin{enumerate}[leftmargin=*, labelindent=0em, itemsep=0.6em]
    \item \textbf{Symmetry}: \(\CL(I_1, I_2) = \CL(I_2, I_1)\) by construction (as shown in Equation \eqref{eq:CL-def}). This property aligns with the core requirement of symmetric similarity measurement in practical interval data analysis.
    
    \item \textbf{Boundedness}: \(\CL(I_1, I_2) \in (0, 1]\). The maximum value 1 is achieved if and only if one interval is fully contained in the other (consistent with the Containment Invariance of one-sided closeness), reflecting complete closeness; the measure approaches 0 as intervals become increasingly separated or dissimilar.
    
    \item \textbf{Vanishing at Extremes}: 
    \(\CL(I_1, I_2) \to 0\) as \(I_2=[c,d]\) moves toward positional infinity or its length approaches 0 or positive infinity. This property ensures intuitive behavior for extreme interval configurations:
    \begin{itemize}[leftmargin=*, itemsep=0.3em]
        \item \textbf{Positional Extremum}: As the midpoint of \(I_2\) (i.e., \(\frac{c+d}{2}\)) moves to positive or negative infinity, the closeness between \(I_1=[a,b]\) and \(I_2=[c,d]\) vanishes:
        \[
        \lim_{\left| \frac{c+d}{2} \right| \to +\infty} \CL(I_1, I_2) = 0
        \]
        \item \textbf{Length Extremum}: When \(I_2=[c,d]\) shrinks to a point (\(|I_2|=d-c \to 0\)) or expands infinitely (\(|I_2|=d-c \to +\infty\)), the weighted closeness to \(I_1=[a,b]\) tends to 0:
        \[
        \lim_{d-c \to 0 \text{ or } +\infty} \CL(I_1, I_2) = 0
        \]
    \end{itemize}
\end{enumerate}

\begin{figure}[ht]
\centering
\includegraphics[width=5cm]{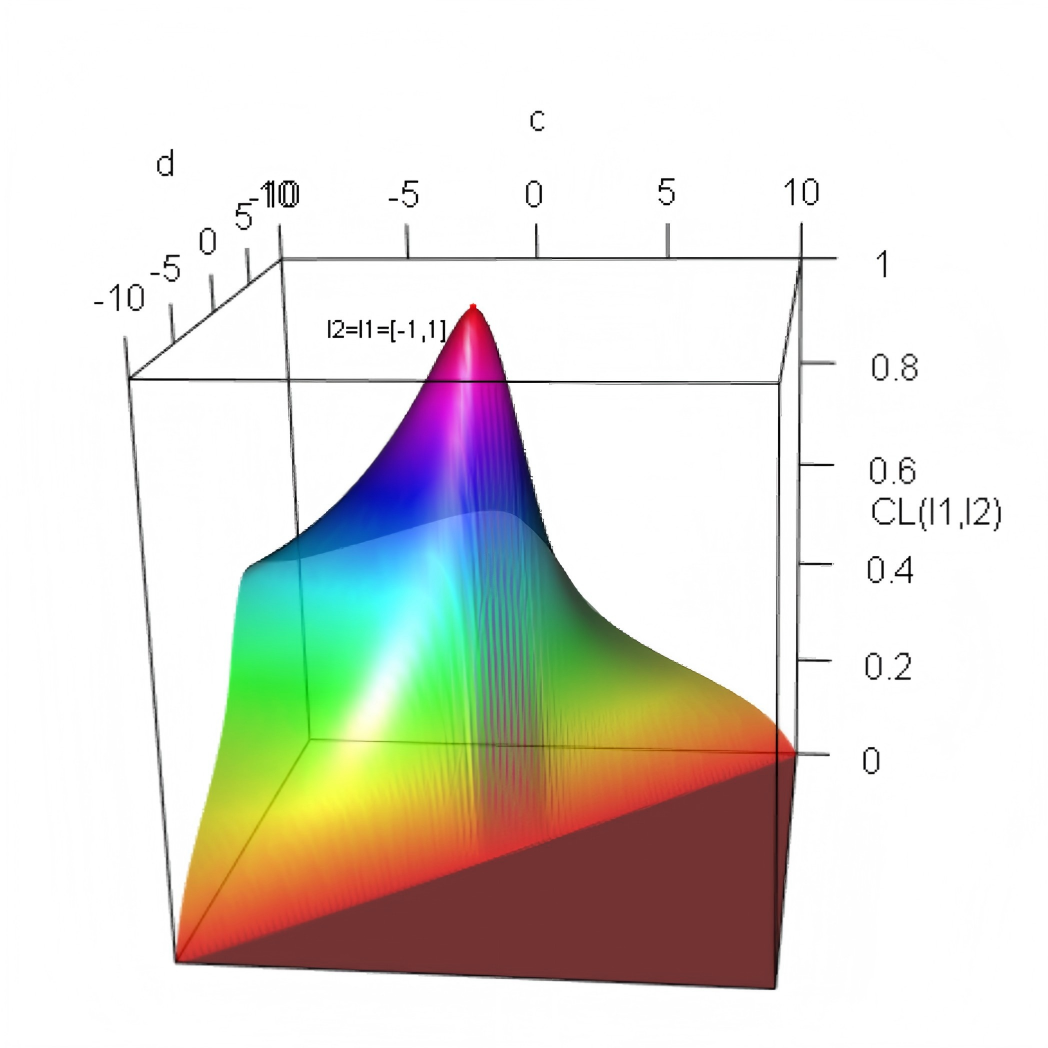}
\caption{3D surface plot of $\CL(I_1,I_2)$ with $I_2=[c,d]$ (lower bound $c$, upper bound $d$) and fixed $I_1=[a,b]=[-1,1]$.
The red marker denotes the position where $I_2=I_1=[-1,1]$ (i.e., $c=-1,d=1$), where $\CL=1$ (the maximum value).
The ordered grid of $(c,d)$ pairs ensures valid surface rendering, and the color gradient (from red/high to blue/low) reflects the magnitude of $\CL$.}
\label{fig:cl_sym_3d}
\end{figure}

To visually validate the aforementioned properties and provide a comprehensive characterization of $\CL$, we construct a 3D surface plot by fixing $I_1=[a,b]=[-1,1]$ (a representative symmetric interval) and treating the lower bound ($c$) and upper bound ($d$) of $I_2=[c,d]$ as bivariate variables. Figure \ref{fig:cl_sym_3d} exhibits the value of $\CL(I_1,I_2)$ across the entire parameter space of $I_2=[c,d]$, with color gradients reflecting the magnitude of $\CL$ (denser colors indicate higher closeness values).

Notably, the 3D surface plot intuitively demonstrates the consistent and predictable behavioral pattern of $\CL$ across all configurations of $I_2=[c,d]$, which confirms the rationality of its theoretical design and provides a clear visual basis for its subsequent application in interval depth computation.

\subsection{Multivariate Interval Closeness}
\label{subsec:multi-closeness}

We extend the one-dimensional closeness measure $\mathrm{CL}$ (defined in Equation \eqref{eq:CL-def}) to multi-dimensional intervals via geometric aggregation, addressing the need for consistent similarity measurement across multi-dimensional hyperrectangles. Let $\mathbf{I}, \mathbf{J} \in \mathcal{I}_d$ denote two $d$-dimensional hyperrectangles, where the $k$-th one-dimensional interval component of $\mathbf{I}$ and $\mathbf{J}$ are given by $[x_k^L, x_k^U]$ and $[y_k^L, y_k^U]$ for all dimensions $k = 1,2,\dots,d$, with $x_k^L \leq x_k^U$ and $y_k^L \leq y_k^U$ holding by definition of an interval.

The multi-dimensional interval closeness integrates per-dimension similarity through geometric averaging, which enforces dimension independence and coordinated alignment across axes.

\begin{definition}[Multivariate Interval Closeness]
\label{def:multi-closeness}
The $d$-dimensional closeness $\mathrm{CL}_{\oplus}(\mathbf{I}, \mathbf{J})$ between two $d$-dimensional hyperrectangles $\mathbf{I}, \mathbf{J} \in \mathcal{I}_d$ is defined as
\[
\mathrm{CL}_{\oplus}(\mathbf{I}, \mathbf{J}) = \left( \prod_{k=1}^{d} \mathrm{CL}\big([x_k^L, x_k^U], [y_k^L, y_k^U]\big) \right)^{1/d},
\]
where $\mathrm{CL}(\cdot, \cdot)$ denotes the one-dimensional symmetric closeness measure in Equation \eqref{eq:CL-def}, and $[x_k^L, x_k^U]$, $[y_k^L, y_k^U]$ are the $k$-th one-dimensional interval components of $\mathbf{I}$ and $\mathbf{J}$, respectively.
\end{definition}

The geometric aggregation scheme preserves the core properties of the one-dimensional $\mathrm{CL}$ and exhibits the following fundamental characteristics tailored to multi-dimensional settings.

\begin{enumerate}
\item \textit{Dimensional Balance.}
Unlike arithmetic averaging (which allows compensation for low similarity in one dimension with high similarity in others), the geometric mean $\mathrm{CL}_{\oplus}$ amplifies sensitivity to low per-dimension closeness values. This design enforces uniform alignment across all dimensions, as poor behavior in any single axis disproportionately reduces the overall closeness score. This is consistent with real-world scenarios where multi-dimensional similarity requires coordination across all features, such as spatial hypercube overlap detection.

\item \textit{Scale Invariance.}
For any positive uniform scaling factor $\alpha > 0$, the measure remains invariant under isotropic resizing of hyperrectangles, eliminating scale-induced biases in similarity assessment.
Formally,
\[
\mathrm{CL}_{\oplus}(\alpha\mathbf{I}, \alpha\mathbf{J}) = \mathrm{CL}_{\oplus}(\mathbf{I}, \mathbf{J}).
\]

\item \textit{Asymptotic Separability (Partition Consistency).}
If there exists any dimension $k$ for which the one-dimensional closeness tends to zero, i.e.,
\[
\mathrm{CL}([x_k^L, x_k^U], [y_k^L, y_k^U]) \to 0,
\]
then the multi-dimensional closeness also collapses to zero:
\[
\mathrm{CL}_{\oplus}(\mathbf{I}, \mathbf{J}) \to 0.
\]
A special case occurs when $\mathrm{CL}^{(k)} = 0$ for some $k$ (complete separability in dimension $k$), which yields $\mathrm{CL}_{\oplus}(\mathbf{I}, \mathbf{J}) = 0$. This serves as a natural generalization of the one-dimensional vanishing property to multi-dimensional settings.
\end{enumerate}

The selection of geometric mean over arithmetic mean is motivated by the assumption of dimension independence: multi-dimensional similarity is only meaningful when all individual dimensions exhibit adequate alignment, rather than relying on partial compensation across axes. This design makes $\mathrm{CL}_{\oplus}$ particularly suitable for tasks requiring strict multi-dimensional coordination, such as hyperrectangle overlap detection in spatial data analysis or feature alignment in multi-dimensional interval datasets.

To generalize the concept of interval statistical depth from the one-dimensional case to the $d$-dimensional case, we build on the multivariate closeness measure $\mathrm{CL}_\oplus$ (Definition~\ref{def:multi-closeness}) and retain the unified interval notation defined above, yielding a depth function that encompasses both one-dimensional and multi-dimensional interval data.

\subsection{Interval Statistical Depth in \( d \)-Dimensional Space}
\label{subsec:d-dim-depth}

We formalize the statistical depth for \( d \)-dimensional interval data by grounding it in the multivariate closeness measure \( \mathrm{CL}_{\oplus} \) (Definition~\ref{def:multi-closeness}), and first define the complete space of \( d \)-dimensional interval objects and its sample subspace for observational data.

The complete space of \( d \)-dimensional intervals is given by
\begin{equation}
\label{eq:d-space}
\mathcal{I}_d
= \left\{
\mathbf{I} = \big([x_1^L,x_1^U],\dots,[x_d^L,x_d^U]\big)
\,\big|\,
x_k^L \leq x_k^U,\; \forall\,k=1,\dots,d
\right\},
\end{equation}
where superscripts \( L \) and \( U \) denote the lower and upper bounds of the \( k \)-th one-dimensional interval component of the \( d \)-dimensional interval vector \( \mathbf{I} \), respectively.
For a finite sample of \( n \) \( d \)-dimensional interval observations, the sample interval subspace is defined as
\begin{equation}
\label{eq:sample-space}
\Omega_n = \big\{ \mathbf{I}_i \big\}_{i=1}^n \subset \mathcal{I}_d,
\end{equation}
where each \( \mathbf{I}_i = ([x_{i1}^L,x_{i1}^U],\dots,[x_{id}^L,x_{id}^U]) \), and \( x_{ik}^L \) and \( x_{ik}^U \) represent the lower and upper bounds of the \( k \)-th dimension for the \( i \)-th sample interval \( \mathbf{I}_i \).

\begin{definition}[Multivariate Interval Depth ($\mathrm{CLD}$)]
\label{def:multidim-depth}

For a target $d$-dimensional interval $\mathbf{I}_0 \in \Omega_n$ and a sample interval subspace $\Omega_n \subset \mathcal{I}_d$, the CL-based interval statistical depth $\mathrm{CLD}(\mathbf{I}_0, \Omega_n)$ is defined as the average of the multivariate closeness between $\mathbf{I}_0$ and all sample intervals in $\Omega_n$:
\begin{equation}
\label{eq:CLD}
\mathrm{CLD}(\mathbf{I}_0, \Omega_n) = \frac{1}{n} \sum_{j=1}^n \mathrm{CL}_{\oplus}(\mathbf{I}_0, \mathbf{I}_j),
\end{equation}
where $\mathrm{CL}_{\oplus}$ is the geometric mean multivariate closeness operator (Definition~\ref{def:multi-closeness}), with its explicit form for sample intervals $\mathbf{I}_p, \mathbf{I}_q \in \Omega_n$ (respectively) given by
\[
\mathrm{CL}_{\oplus}(\mathbf{I}_p, \mathbf{I}_q) = \left( \prod_{k=1}^d \mathrm{CL}\big( [x_{pk}^L, x_{pk}^U], [x_{qk}^L, x_{qk}^U] \big) \right)^{1/d},
\]
and $\mathrm{CL}(\cdot, \cdot)$ denotes the one-dimensional symmetric closeness measure (Equation~\eqref{eq:CL-def}).
\end{definition}

The CLD statistic satisfies the following fundamental properties of statistical depth functions.

\begin{enumerate}
\item \textit{Vanishing at Infinity.}
For any interval \( \mathbf{I} \in \mathcal{I}_d \), as \( \mathbf{I} \) moves infinitely far from the sample cloud \( \Omega_n \) in at least one dimension, the one-dimensional closeness \( \mathrm{CL} \) between the \( k \)-th component of \( \mathbf{I} \) and that of each sample interval \( \mathbf{I}_j \) tends to 0.
As the geometric mean of dimension-wise \( \mathrm{CL} \) values, \( \mathrm{CL}_\oplus(\mathbf{I}, \mathbf{I}_j) \) also tends to 0 for all \( \mathbf{I}_j \in \Omega_n \).
Thus, the sample-averaged \( \mathrm{CLD} \) converges to zero, i.e.,
\[
\lim_{\mathrm{dist}(\mathbf{I}, \Omega_n) \to +\infty} \mathrm{CLD}(\mathbf{I}, \Omega_n) = 0,
\]
where \( \mathrm{dist}(\mathbf{I}, \Omega_n) \) denotes the standard set distance between \( \mathbf{I} \) and \( \Omega_n \) across all dimensions.

\item \textit{Affine Invariance (Translation and Uniform Scaling).}
The one-dimensional closeness \( \mathrm{CL} \) is invariant to global translation and uniform scaling.
The measure \( \mathrm{CL}_\oplus \) inherits this invariance via the geometric mean, and \( \mathrm{CLD} \) preserves it as a sample average.
This ensures scale- and location-free inference for multivariate interval data.
Formally, for any affine transformation \( A \) on \( \mathcal{I}_d \) composed of translation and uniform scaling,
\[
\mathrm{CLD}(A(\mathbf{I}), A(\Omega_n)) = \mathrm{CLD}(\mathbf{I}, \Omega_n).
\]
\end{enumerate}

The properties of \( \mathrm{CL} \), \( \mathrm{CL}_\oplus \), and the two core properties of \( \mathrm{CLD} \) stated above are rigorously proven in Appendix \ref{app:proofs-cld}.

\section{SIMULATIONS}
To evaluate the effectiveness and practical applicability of the proposed closeness depth framework, we conduct a series of simulation experiments. We generate two types of synthetic interval-valued datasets: one in a univariate (1D) setting and another in a multivariate (3D) setting. These simulations are designed to illustrate the robustness, interpretability, and anomaly detection capability of the proposed depth measure across different dimensionalities.

Detailed results and visualizations for both univariate and multivariate scenarios are presented in the following sections.

\subsection{One-Dimensional Interval Data Simulation}
\label{1d}
We perform a simulation study to compare the anomaly detection performance of Closeness Depth with three classical depth measures: Spatial Depth \cite{Vardi2000}, Half-space Depth \cite{Tukey1975}, and Mahalanobis Depth \cite{Liu1993}. Experiments are conducted under three types of anomalies—position anomalies, length anomalies, and composite anomalies—along with varying levels of contamination.

\subsubsection{Data Settings}

Consider a univariate interval dataset with $n$ samples, where the $i$-th interval is denoted $I_i = [x_i^L, x_i^U]$ for $i = 1, 2, \dots, n$.
Each interval is parameterized by its center $c_i$ and length $l_i > 0$.
The lower and upper bounds are then given by
\[
x_i^L = c_i - \frac{l_i}{2}, \quad x_i^U = c_i + \frac{l_i}{2}.
\]
Using this parameterization, we generate normal and anomalous intervals according to the following distributions:
\begin{itemize}
    \item Normal intervals: $c_i \sim \mathcal{N}(50, 15^2)$, $l_i \sim \mathcal{U}(10, 20)$;
    \item Type I (Position anomalies): $c_i \sim \mathcal{U}(0,25)$ or $\mathcal{U}(75,100)$ (each with probability $0.5$), $l_i \sim \mathcal{U}(10,20)$;
    \item Type II (Length anomalies): $c_i \sim \mathcal{N}(50, 15^2)$, $l_i \sim \mathcal{U}(0,5)$;
    \item Type III (Composite anomalies): $c_i$ follows Type I, $l_i$ follows Type II.
\end{itemize}

In our experiments, we set the number of normal samples to $n = 200$,
and consider three contamination ratios $\delta \in \{0.1, 0.15, 0.2\}$.
Each dataset contains $\lfloor \delta n \rfloor$ anomalous samples for each anomaly type.

\begin{figure}[ht]
\centering
\begin{subfigure}[t]{0.32\linewidth}
    \centering
    \includegraphics[width=\linewidth]{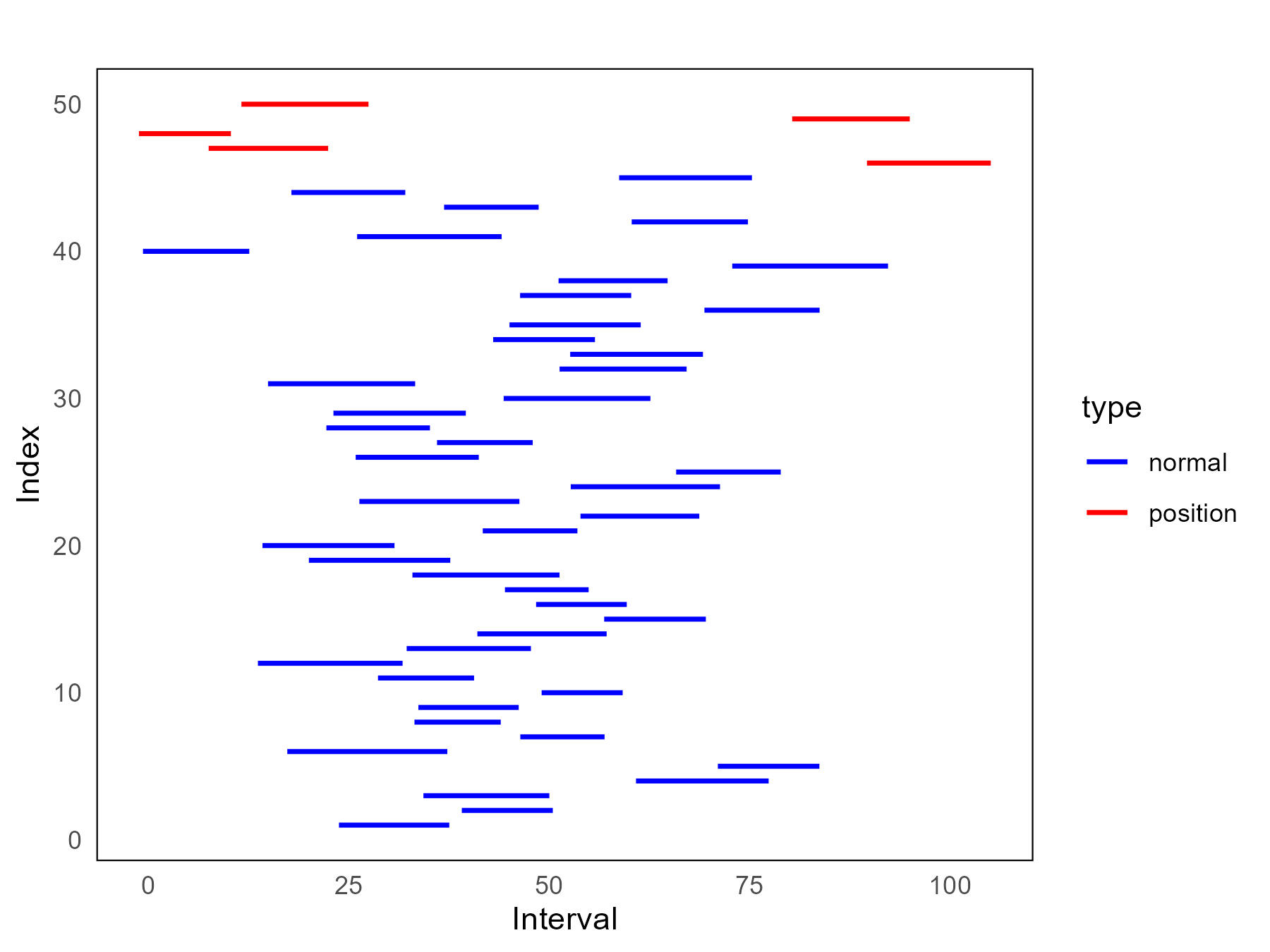}
    \caption{Position anomalies}
\end{subfigure}
\hfill
\begin{subfigure}[t]{0.32\linewidth}
    \centering
    \includegraphics[width=\linewidth]{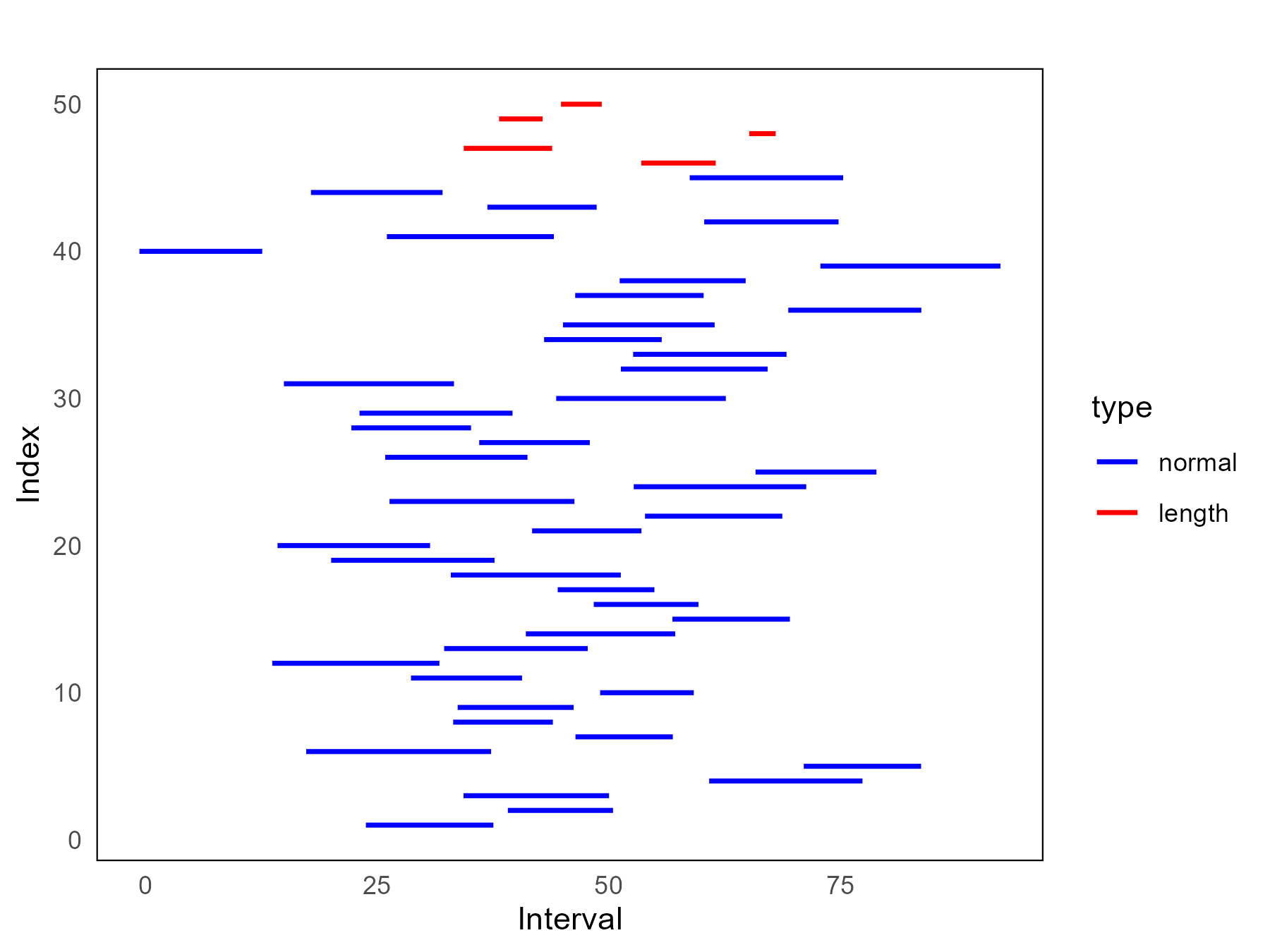}
    \caption{Length anomalies}
\end{subfigure}
\hfill
\begin{subfigure}[t]{0.32\linewidth}
    \centering
    \includegraphics[width=\linewidth]{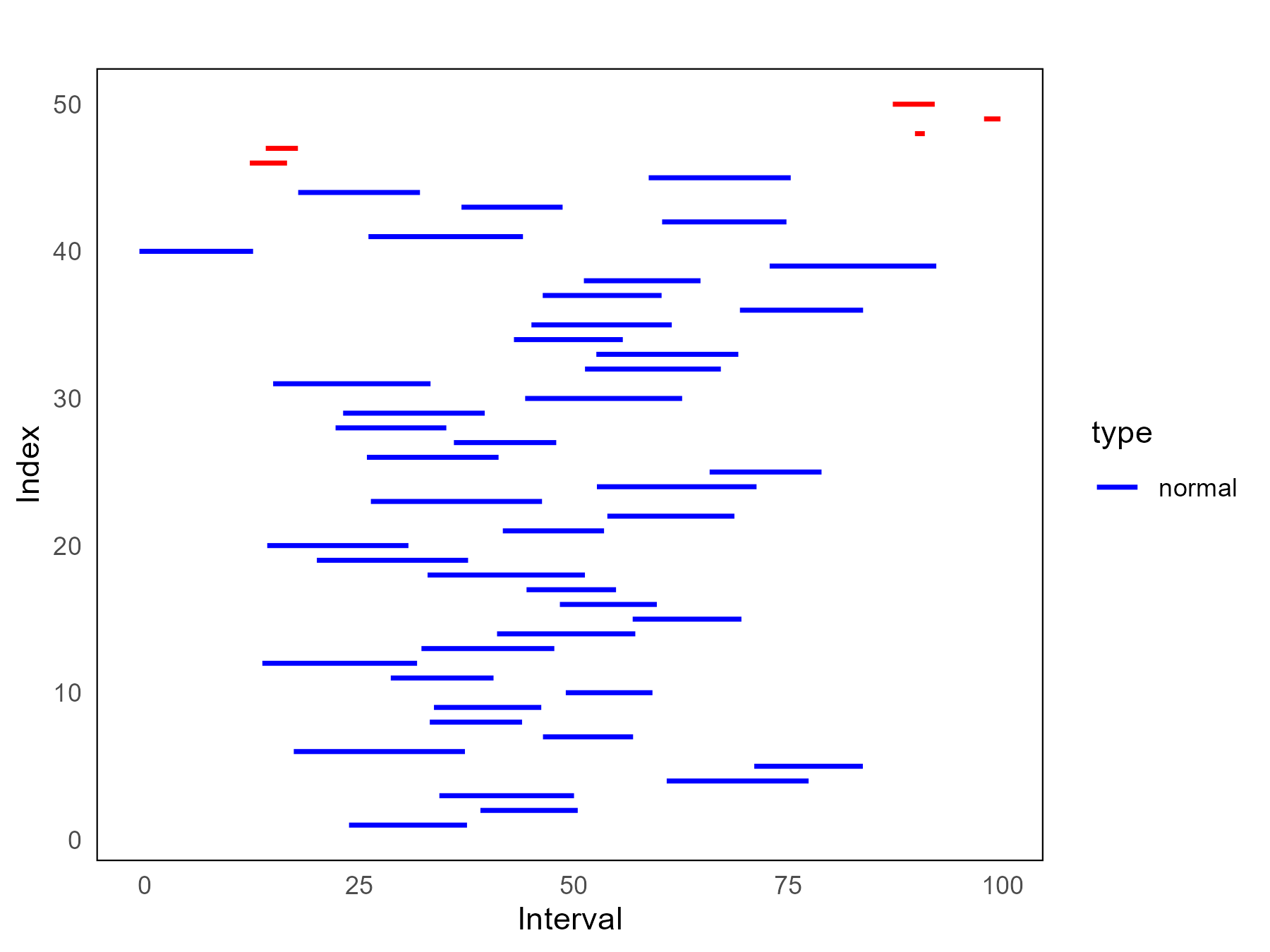}
    \caption{Composite anomalies}
\end{subfigure}
\caption{Visualization of different anomaly categories for univariate interval data}\label{fig:combo}
\end{figure}

Visualizations of the three anomaly categories (with $n=50$ and $\delta=0.1$ for illustration) are shown in Figure \ref{fig:combo}.

\subsubsection{Depth Calculation}

In this simulation study, we employ four statistical depth measures:
the proposed Closeness Depth (CLD) introduced in Section \ref{subsec:d-dim-depth},
and three classical methods: Spatial Depth, Half-Space Depth, and Mahalanobis Depth.
All classical approaches represent each univariate interval as a 2-dimensional vector composed of its lower and upper bounds.

Since the center-length parameterization is linearly invertible to the lower-upper bound representation,
and all considered depth measures are invariant under invertible linear transformations,
the depth ranking used for anomaly detection remains consistent across both representations.

\subsubsection{Simulation Procedure}

The standard depth-based anomaly detection pipeline is as follows:
(1) Compute the depth value of each interval using each measure;
(2) Since lower depth indicates greater anomalousness, select the $\lfloor \delta n \rfloor$ intervals with the smallest depth values as detected anomalies;
(3) Evaluate performance using two metrics: Precision (proportion of true anomalies among detected anomalies) and False Positive Rate (FPR, proportion of normal samples incorrectly identified as anomalies).

To ensure statistical robustness, each experimental setting is repeated 50 times with independent random seeds.
The results reported in the tables represent the mean and standard deviation (in parentheses) over the 50 independent trials.

The anomaly detection performance under the three contamination ratios is summarized as follows:

\begin{table*}[ht]
\centering
\caption{Anomaly Detection Performance of Statistical Depth Measures ($n=200, \delta=0.1$)}
\label{tab:3model4method}
\small
\begin{tabular}{l *{8}{c}}
\toprule
\multirow{2}{*}{Anomaly Type} & \multicolumn{8}{c}{Statistical Depth Measures} \\
\cmidrule(lr){2-9}
& \multicolumn{2}{c}{Closeness} & \multicolumn{2}{c}{Spatial} & \multicolumn{2}{c}{Half-space} & \multicolumn{2}{c}{Mahalanobis} \\
\cmidrule(lr){2-3} \cmidrule(lr){4-5} \cmidrule(lr){6-7} \cmidrule(lr){8-9}
& Prec. (\%) & FPR (\%) & Prec. (\%) & FPR (\%) & Prec. (\%) & FPR (\%) & Prec. (\%) & FPR (\%) \\
\midrule
Position Anomaly & 77.8 (7.1) & 2.2 & 68.4 (8.0) & 3.2 & 44.3 (9.9) & 5.6 & 72.9 (7.2) & 2.7 \\
Length Anomaly & 70.3 (6.4) & 3.0 & 54.5 (5.7) & 4.6 & 32.0 (5.5) & 6.8 & 77.4 (5.9) & 2.3 \\
Composite Anomaly & 99.5 (1.5) & 0.0 & 84.8 (5.0) & 1.5 & 40.1 (5.8) & 6.0 & 99.2 (0.3) & 0.1 \\
\bottomrule
\end{tabular}
\end{table*}

\begin{table*}[ht]
\centering
\caption{Anomaly Detection Performance of Statistical Depth Measures ($n=200, \delta=0.15$)}
\label{tab:delta15_corrected}
\small
\begin{tabular}{l *{8}{c}}
\toprule
\multirow{2}{*}{Anomaly Type} & \multicolumn{8}{c}{Statistical Depth Measures} \\
\cmidrule(lr){2-9}
& \multicolumn{2}{c}{Closeness} & \multicolumn{2}{c}{Spatial} & \multicolumn{2}{c}{Half-space} & \multicolumn{2}{c}{Mahalanobis} \\
\cmidrule(lr){2-3} \cmidrule(lr){4-5} \cmidrule(lr){6-7} \cmidrule(lr){8-9}
& Prec. (\%) & FPR (\%) & Prec. (\%) & FPR (\%) & Prec. (\%) & FPR (\%) & Prec. (\%) & FPR (\%) \\
\midrule
Position Anomaly & 81.9 (5.0) & 2.7 & 69.5 (6.2) & 4.6 & 51.2 (5.7) & 7.3 & 75.5 (5.8) & 3.7 \\
Length Anomaly & 72.5 (6.4) & 4.1 & 50.2 (4.2) & 7.5 & 35.9 (3.9) & 9.6 & 70.9 (5.3) & 4.4 \\
Composite Anomaly & 99.6 (1.1) & 0.1 & 80.9 (3.8) & 2.9 & 44.8 (4.1) & 8.3 & 98.7 (2.0) & 0.2 \\
\bottomrule
\end{tabular}
\end{table*}

\begin{table*}[ht]
\centering
\caption{Anomaly Detection Performance of Statistical Depth Measures ($n=200, \delta=0.2$)}
\label{tab:delta20_results}
\small
\begin{tabular}{l *{8}{c}}
\toprule
\multirow{2}{*}{Anomaly Type} & \multicolumn{8}{c}{Statistical Depth Measures} \\
\cmidrule(lr){2-9}
& \multicolumn{2}{c}{Closeness} & \multicolumn{2}{c}{Spatial} & \multicolumn{2}{c}{Half-space} & \multicolumn{2}{c}{Mahalanobis} \\
\cmidrule(lr){2-3} \cmidrule(lr){4-5} \cmidrule(lr){6-7} \cmidrule(lr){8-9}
& Prec. (\%) & FPR (\%) & Prec. (\%) & FPR (\%) & Prec. (\%) & FPR (\%) & Prec. (\%) & FPR (\%) \\
\midrule
Position Anomaly & 84.5 (3.9) & 3.1 & 69.2 (5.6) & 6.2 & 55.4 (5.2) & 8.9 & 75.4 (5.1) & 4.9 \\
Length Anomaly & 74.8 (4.5) & 5.0 & 48.7 (4.5) & 10.3 & 36.9 (3.6) & 12.6 & 66.1 (4.9) & 6.8 \\
Composite Anomaly & 99.8 (0.9) & 0.0 & 77.8 (3.3) & 4.4 & 46.6 (3.7) & 10.7 & 98.7 (1.8) & 0.3 \\
\bottomrule
\end{tabular}
\end{table*}

\subsubsection{Results Analysis}

As reflected in the numerical results (Tables \ref{tab:3model4method}, \ref{tab:delta15_corrected}, and \ref{tab:delta20_results}), the proposed Closeness Depth (CLD) exhibits strong discriminatory power for position anomalies, short-length anomalies, and composite anomalies with simultaneous position and length deviations. Across all contamination ratios ($\delta=0.1, 0.15, 0.2$) and most anomaly types, CLD outperforms the three classical depth measures. The only exception occurs at $\delta=0.1$, where Mahalanobis Depth achieves slightly higher precision than CLD for short-length anomalies. This advantage of CLD stems from its ability to accurately quantify the centrality of interval data by aggregating pairwise closeness between intervals, enabling effective identification of anomalous intervals as samples with low depth values. It performs excellently in detecting the three anomaly types, and particularly captures the unique geometric feature of short-length intervals with narrow ranges to achieve clear differentiation from normal intervals.

Notably, CLD demonstrates varying performance across different anomaly types. It achieves nearly perfect detection accuracy for composite anomalies, with precision close to 100\%. The detection accuracy for position anomalies gradually improves with increasing contamination ratio, rising from 77.8\% to 84.5\%. Although its accuracy for short-length anomalies is generally superior to that of Spatial Depth and Half-Space Depth, it is slightly inferior to Mahalanobis Depth at $\delta=0.1$.

This performance variation is essentially related to the geometric characteristics of anomaly types and the computational mechanism of CLD. Composite anomalies possess dual features of position offset and length abnormality, resulting in extremely low closeness to normal intervals and thus being easily identified as low-depth samples. Position anomalies have similar interval ranges to normal samples but significant central offsets, and CLD can capture such offset differences through pairwise closeness aggregation. With a higher contamination ratio, the anomaly signal becomes more concentrated, leading to improved detection accuracy. Short-length anomalies only exhibit the feature of narrow ranges, and some such samples may maintain a certain degree of closeness to the edges of normal intervals, allowing Mahalanobis Depth to gain a slight advantage under low contamination ratios.

Therefore, when interpreting depth values and applying them to anomaly detection in practice, targeted analysis combined with the characteristics of specific anomaly types is recommended. CLD shows strong applicability in most real-world scenarios due to its effective discrimination for position offsets and composite anomalies. Meanwhile, its stably low False Positive Rate (FPR), all below 5\%, reflects reliable practical performance and can effectively reduce misclassification of normal intervals. This interpretation framework that integrates anomaly type characteristics and depth values provides a more accurate analytical basis for anomaly detection using interval data.

\subsection{Three-Dimensional Cube Anomaly Detection Simulation}

In many real-world applications, data are inherently multidimensional, and three-dimensional (3D) spatial representations are particularly common, ranging from medical imaging to industrial quality control. While one-dimensional interval data simulations provide useful insights, extending anomaly detection to 3D space is essential for practical applications, as it involves unique computational challenges such as the curse of dimensionality and the need for geometry-aware anomaly scoring.

This simulation study investigates the effectiveness of closeness depth-based anomaly detection for 3D cubes. In contrast to traditional depth metrics such as Mahalanobis depth, the proposed method explicitly models both positional deviations and volumetric anomalies within a unified depth framework. By representing each cube in terms of its spatial coordinates and edge lengths, we compute depth values that reflect the relative normality of each sample within the dataset.

\subsubsection{Data Setting}

This section constructs a 3D cube interval dataset to evaluate the anomaly detection performance of the proposed Closeness Depth (CLD) and two comparative depth methods.

Let the center coordinates of a 3D cube sample be $(x_c, y_c, z_c)$, and the edge lengths along the three axes be $l_x, l_y, l_z$.
The interval bounds in each dimension are defined as
\[
x^L = \max\left(x_c - \frac{l_x}{2}, 0\right),\quad x^U = \min\left(x_c + \frac{l_x}{2}, 100\right),
\]
\[
y^L = \max\left(y_c - \frac{l_y}{2}, 0\right),\quad y^U = \min\left(y_c + \frac{l_y}{2}, 100\right),
\]
\[
z^L = \max\left(z_c - \frac{l_z}{2}, 0\right),\quad z^U = \min\left(z_c + \frac{l_z}{2}, 100\right),
\]
where the max and min functions act as boundary clipping to ensure all samples lie within the 3D region $[0,100]^3$.
Each sample is represented as a 3D interval cube, denoted by
\[
\mathcal{X} = [x^L, x^U] \times [y^L, y^U] \times [z^L, z^U].
\]

Normal samples are generated from uniform distributions to concentrate in the core region with consistent geometric morphology.
The center coordinates and edge lengths are independently and identically distributed as
\[
x_c, y_c, z_c \stackrel{\text{i.i.d.}}{\sim} \mathcal{U}(20, 80),\quad l_x, l_y, l_z \stackrel{\text{i.i.d.}}{\sim} \mathcal{U}(10, 20).
\]
A total of $N_0$ normal samples are generated independently.

Anomalous samples are divided into three types: position anomalies, size anomalies, and composite anomalies.
Each type is randomly sampled to simulate realistic abnormal patterns.
A total of $N_1$ anomalous samples are generated:
\begin{itemize}
    \item Type I (Position anomalies): $x_c, y_c, z_c \sim \mathcal{U}(0,30)$ or $\mathcal{U}(70,100)$ (each with probability $0.5$), $l_x, l_y, l_z \sim \mathcal{U}(10,20)$;
    \item Type II (Size anomalies): $x_c, y_c, z_c \sim \mathcal{U}(20,80)$, $l_x, l_y, l_z \sim \mathcal{U}(1,9)$ or $\mathcal{U}(30,50)$ (each with probability $0.5$);
    \item Type III (Composite anomalies): $x_c, y_c, z_c$ follow the rule of Type I, and $l_x, l_y, l_z$ follow the rule of Type II.
\end{itemize}

In this simulation, we set $N_0=200$ and $N_1=15$.

\subsubsection{Depth Calculation}
Three depth measures are employed for cube ranking: Closeness Depth, Spatial Depth, and Mahalanobis Depth.
Spatial Depth and Mahalanobis Depth are used as in the univariate setting (Section \ref{1d}), and are directly extended to the 3D case by treating each cube as a high-dimensional vector constructed from its interval bounds.
Since these methods were not originally designed for interval-valued data, no special adaptation is performed beyond this straightforward vectorization.

In contrast, the proposed Closeness Depth is specifically constructed for 3D interval data, enabling it to better capture both the spatial location and size characteristics of cubes.
Depth values are used to identify outlying cubes that deviate from the majority of samples, where a lower depth value indicates a higher likelihood of being anomalous.

\subsubsection{Simulation Procedure}
To ensure the reliability and statistical significance of the experimental results, the anomaly detection simulation is repeated 50 independent times under the same experimental settings.

In each repetition, the three depth measures are applied to compute the depth value for each cube.The cubes are then ranked in ascending order of depth, and the top $N_1$ cubes are selected as detected anomalies.
Precision is defined as the percentage of true anomalies among detected anomalies, and its mean and standard deviation over 50 independent runs are reported in Table \ref{tab:simulation_50times}.

\begin{table}[htbp]
    \centering
    \caption{Precision Performance of Three Depth Measures}
    \label{tab:simulation_50times}
    \fontsize{8}{9.6}\selectfont
    \setlength{\tabcolsep}{10pt}
    \begin{tabular}{@{}l cc @{}}
        \toprule
        Method & Precision & Standard Deviation \\
        \midrule
        CLD Depth & 0.864 & 0.1432 \\
        Spatial Depth & 0.648 & 0.1502 \\
        Mahalanobis Depth & 0.634 & 0.1287 \\
        \bottomrule
    \end{tabular}
\end{table}
CLD Depth achieves the highest precision (0.864) with a standard deviation of 0.1432, outperforming both Spatial Depth (0.648, SD = 0.1502) and Mahalanobis Depth (0.634, SD = 0.1287) across all anomaly types, demonstrating stronger overall robustness and stability which arises from its ability to integrate both positional and volumetric information. Traditional depth measures exhibit clear limitations in this task: Spatial Depth relies on geometric centrality assumptions, leading to relatively low detection precision, while Mahalanobis Depth is sensitive to the sample covariance matrix, limiting its adaptability to the complex distributional structure of 3D interval data.

These results confirm that CLD Depth serves as a more reliable anomaly detection tool, especially in scenarios where classical depth measures are less effective.
Mahalanobis Depth and other parametric approaches depend on elliptical or Gaussian distribution assumptions, which can degrade performance under non-standard or contaminated data geometries.
In contrast, CLD Depth is distribution-free and does not require any parametric assumptions, allowing it to adaptively identify anomalies by measuring pairwise closeness in both location and scale.
This dual characterization ensures consistent performance across position anomalies, size anomalies, and composite anomalies, while avoiding the restrictive assumptions inherent to conventional depth methods.

The simulation results are visualized in 3D plots.
For clarity, a subset of normal cubes is shown in blue, all true anomalies in red, and the top detected anomalies with the smallest depth values are highlighted in green.
\begin{figure}[ht]
\centering
\includegraphics[width=5cm]{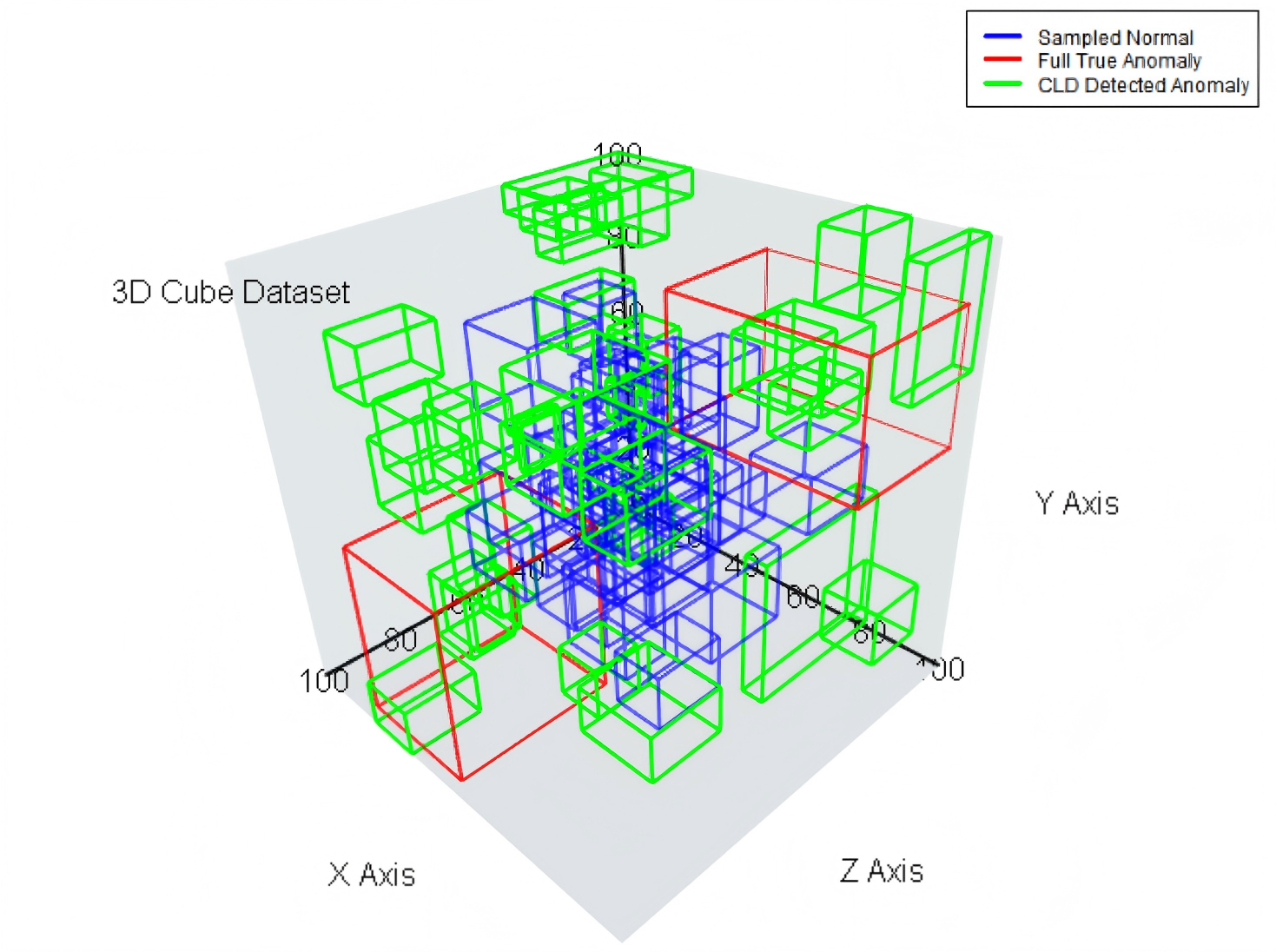}
\caption{3D visualizations of the anomalies}\label{fig:3dinterval}
\end{figure}

\subsubsection{Results and Analysis}

From the visualization examples in Figure \ref{fig:3dinterval}, combined with the quantitative results in Table \ref{tab:simulation_50times},
we further analyze the behavior of depth-based anomaly detection in 3D cube data.

Overall, the depth-based approach effectively distinguishes anomalies from normal cubes,
as evident from the fact that most anomalous samples are assigned relatively low depth values.
Notably, cubes with large positional deviations tend to exhibit extremely low depth and can be readily detected.
In comparison, cubes with only size deviations are relatively more difficult to identify,
since they maintain considerable spatial overlap with normal cubes and thus receive higher depth values.

These observations illustrate how geometric characteristics of anomalies
affect depth values and further explain the superior performance of CLD Depth,
which jointly considers both positional and volumetric information when measuring the relative centrality of 3D intervals.

\section{REAL DATA EXPERIMENT}

In this section, we apply the proposed depth-based anomaly detection approach to two real-world datasets from different domains: stock market data and GPS trajectory data.
These datasets reflect typical scenarios in financial analysis and human mobility mining, and are used to validate the effectiveness of the depth-based method in identifying practical anomalies.

The stock market experiment is conducted on Chinese A-share data, focusing on several major bank stocks over a five-month period.
Daily trading prices are represented as intervals formed by daily high and low prices, and the depth-based method is used to detect anomalous price movements and unusual market behaviors.

The GPS trajectory experiment uses a subset of the GeoLife dataset collected in Beijing, which records individual outdoor movements.
Each trajectory is represented by interval-valued latitude and longitude ranges, and the depth-based approach is applied to identify anomalous trajectories that deviate from normal mobility patterns, such as long-distance trips and unusual detours.

Experimental results on both real datasets demonstrate that the depth-based method can effectively detect meaningful anomalies in practice,
and provide reliable support for financial data analysis and abnormal trajectory recognition.

\subsection{Stock Data Anomaly Detection}

\subsubsection{Dataset Overview}
The dataset is sourced from the Chinese A-share market and includes daily trading records in 2024.
Each entry contains Stock Code, Stock Name, Date, Open Price, Highest Price, Lowest Price, Closing Price, Previous Close, Price Change, Price Change Percentage, Volume, and Turnover.
The daily highest and lowest prices naturally form an interval, which is well-suited for the proposed CLD Depth-based anomaly detection method.

\subsubsection{Data Selection and Transformation}
\begin{figure}[ht]
\centering
\includegraphics[width=5cm]{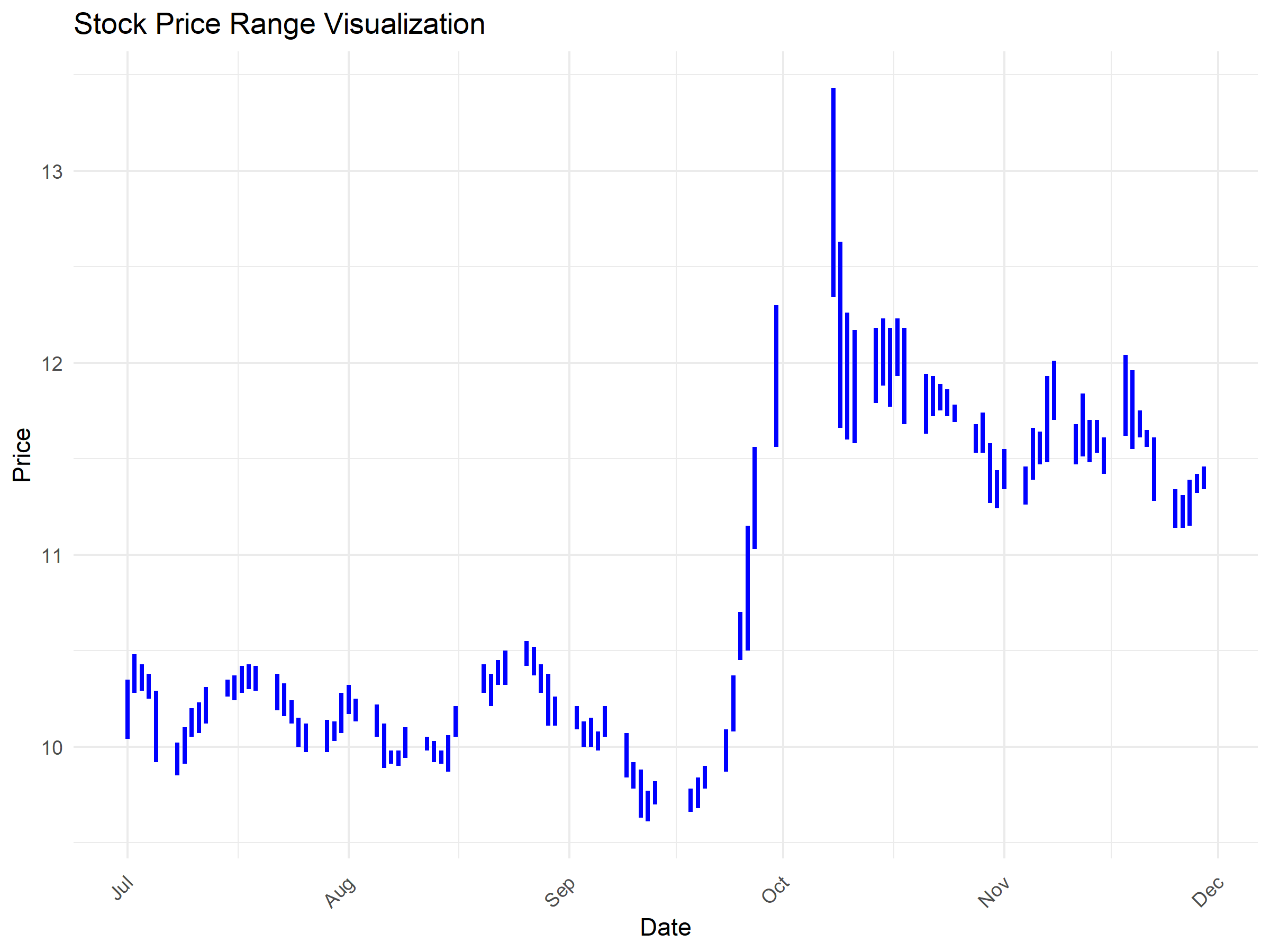}
\caption{Visualization of Stock Price Intervals for \texttt{Ping An Bank} (July 1, 2024 -- November 29, 2024)}
\label{fig:stock_pingan}
\end{figure}

As a representative case, we select stock \texttt{000001} (Ping An Bank) from July 1 to November 29, 2024, covering 102 trading days.
The interval-valued data are constructed using the daily highest price as the upper bound and the lowest price as the lower bound.
The raw interval sequence is visualized in Figure \ref{fig:stock_pingan}.

\subsubsection{Depth Calculation and Anomaly Detection}
\begin{figure}[ht]
\centering
\includegraphics[width=5cm]{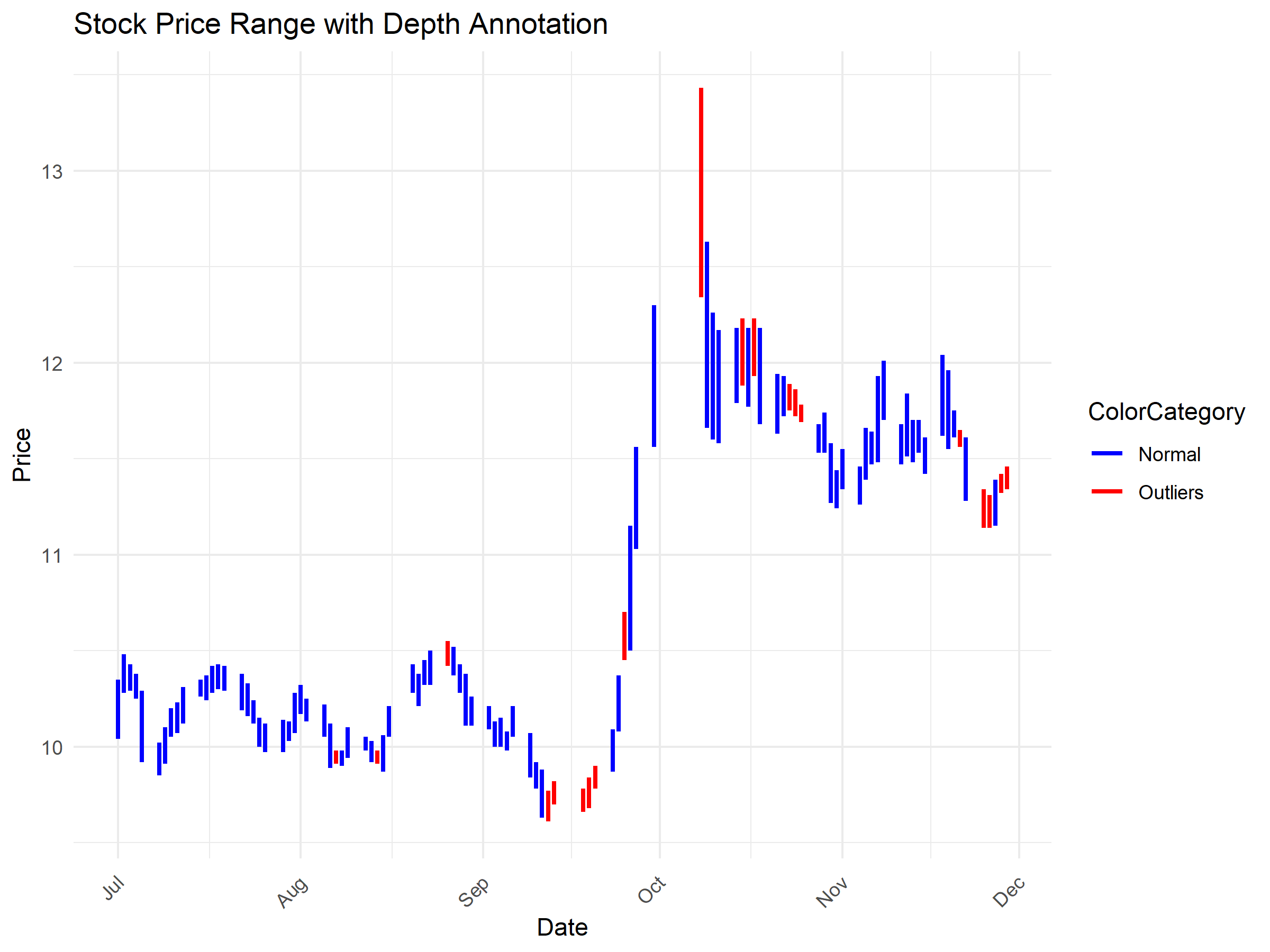}
\caption{Stock Price Intervals with Anomalous Days Highlighted (Red)}
\label{fig:stock_anomaly}
\end{figure}

We compute the CLD value for each daily interval using the method proposed in Section 3.
The depth value measures the relative centrality of each interval within the dataset.
The 20 intervals with the smallest depth values are identified as anomalous, as they deviate most from normal trading patterns.
Figure \ref{fig:stock_anomaly} highlights these anomalous days in red.

\subsubsection{Results and Analysis}
The identified anomalous days generally coincide with sharp price fluctuations and high volatility,
which often correspond to major market events, earnings releases, or external news impacts.
These anomalies typically exhibit:
\begin{itemize}
    \item Large price swings within a single trading day;
    \item Abnormal volatility that deviates from typical trading patterns;
    \item Market responses to important financial events or policy news.
\end{itemize}

\begin{figure}[ht]
\centering
\begin{subfigure}[t]{0.48\linewidth}
    \centering
    \includegraphics[width=\linewidth]{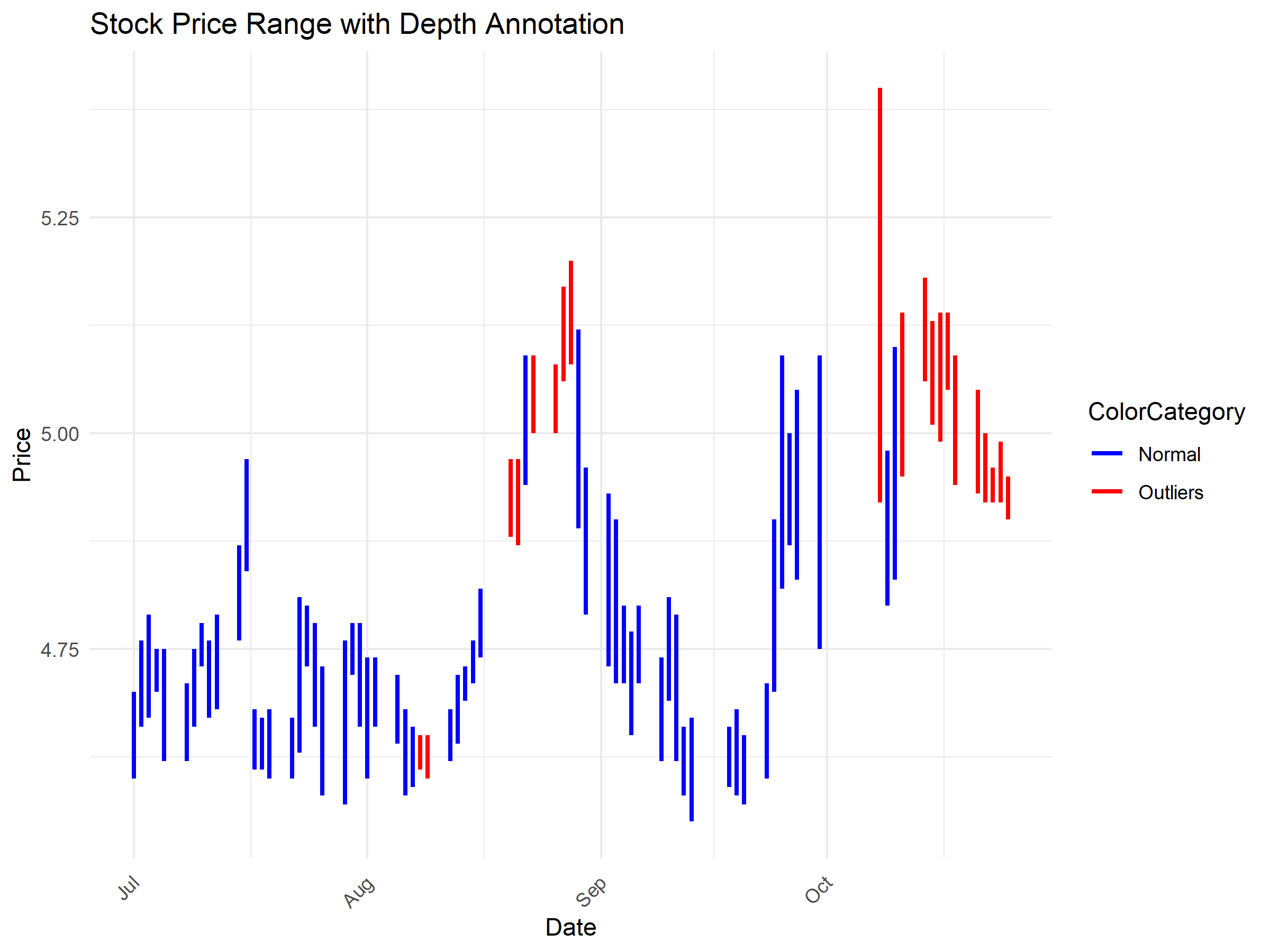}
    \caption{Bank Of China (601988)}
\end{subfigure}
\hfill
\begin{subfigure}[t]{0.48\linewidth}
    \centering
    \includegraphics[width=\linewidth]{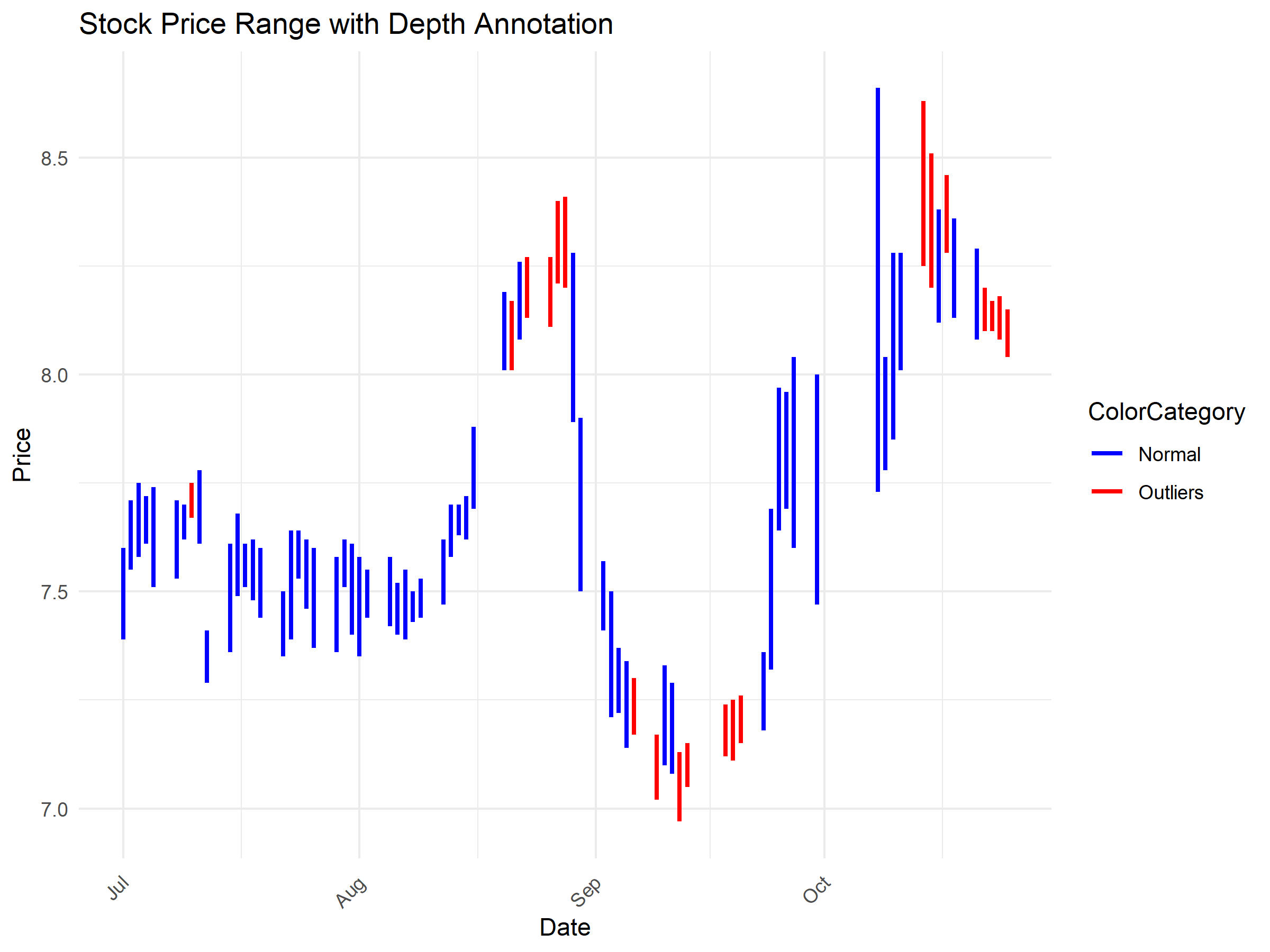}
    \caption{China Construction Bank (601939)}
\end{subfigure}

\vspace{10pt}

\begin{subfigure}[t]{0.48\linewidth}
    \centering
    \includegraphics[width=\linewidth]{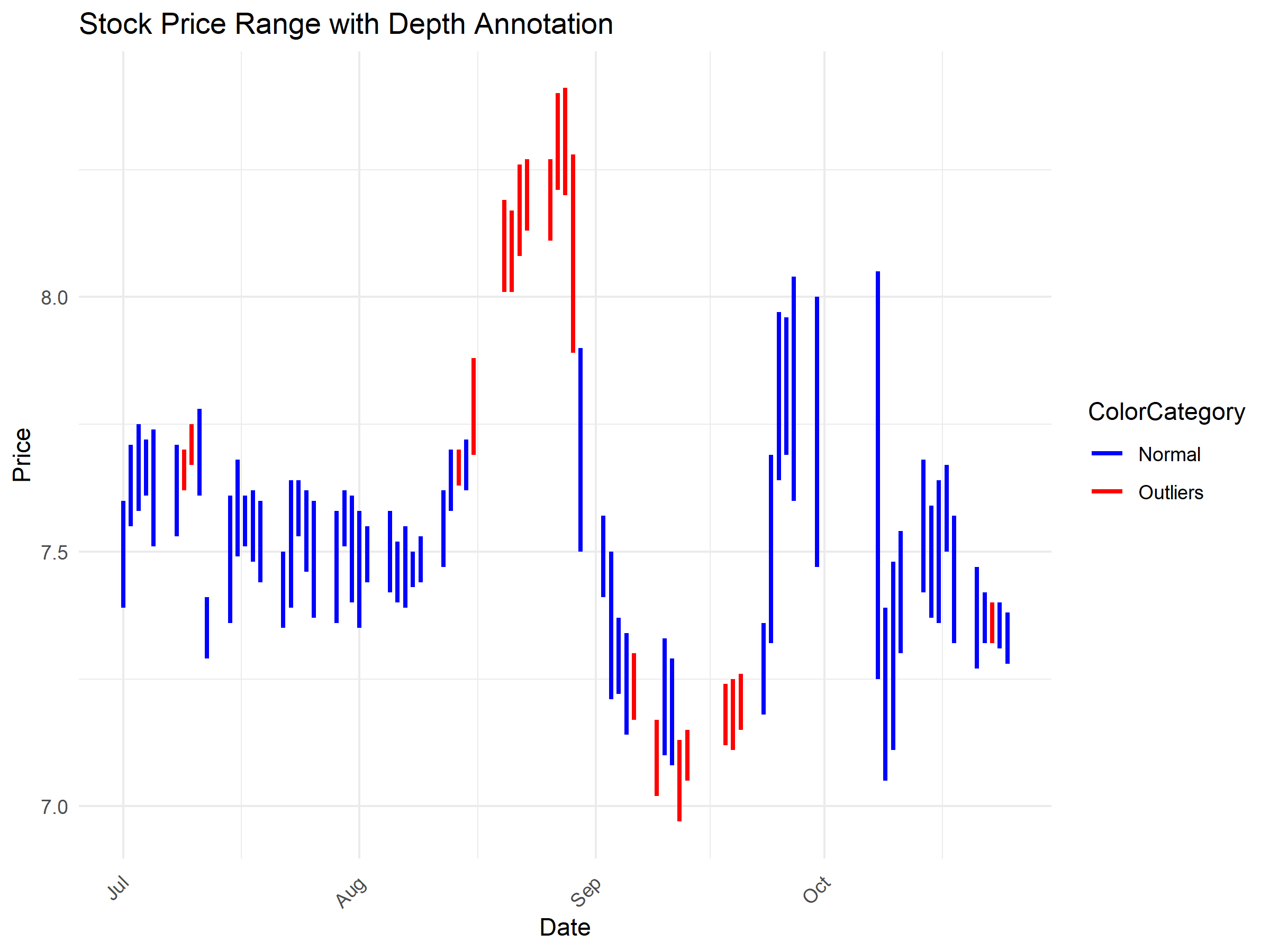}
    \caption{Bank of Communications (601328)}
\end{subfigure}
\hfill
\begin{subfigure}[t]{0.48\linewidth}
    \centering
    \includegraphics[width=\linewidth]{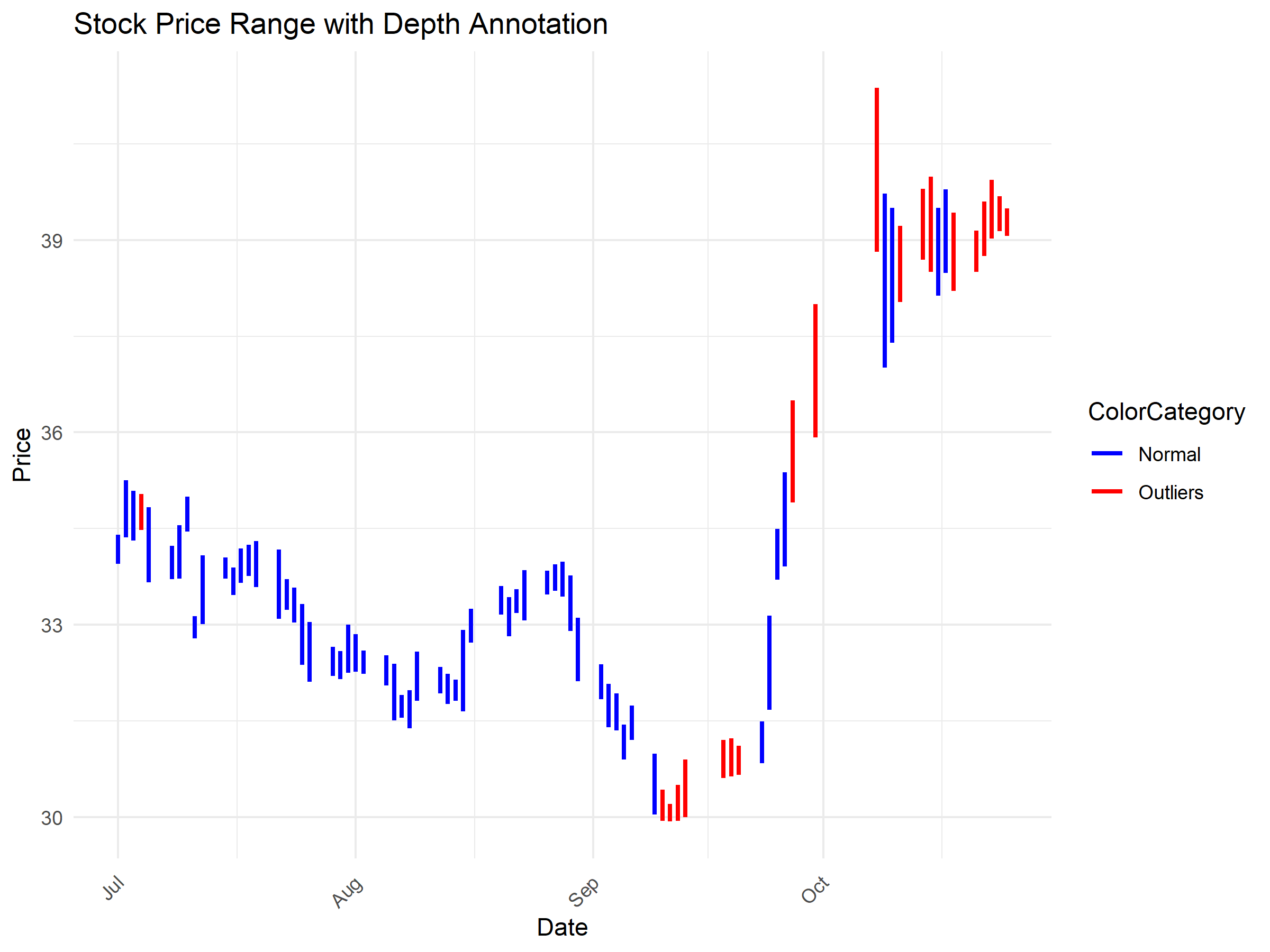}
    \caption{China Merchants Bank (600036)}
\end{subfigure}
\caption{Stock Price Range Visualizations for Four Banks}
\label{fig:bank_stocks}
\end{figure}

To verify the generality of the proposed method, we further apply the same procedure to four other major bank stocks:
\texttt{Bank Of China} (601988), \texttt{China Construction Bank} (601939), \texttt{Bank of Communications} (601328), and \texttt{China Merchants Bank} (600036).
As shown in Figure \ref{fig:bank_stocks}, the method consistently detects anomalous days marked by substantial price movements across all stocks.

Overall, the depth-based approach effectively detects anomalous trading days in real stock market data,
showing strong practical value for financial anomaly detection and market monitoring.

\subsection{ Multivariate Stock Depth Analysis}

In this section, we extend the analysis by considering the daily stock data of five banks as a joint five-dimensional interval dataset. Each dimension corresponds to one of the selected banks. By calculating the depth of each joint interval across the five stocks, we aim to observe the collective market behavior and identify anomalies that indicate unusual market-wide fluctuations.

\subsubsection{Data Transformation}
For this experiment, the stock data of five banks, namely \texttt{Ping An Bank (000001.SZ)}, \texttt{Bank of China (601988)}, \texttt{China Construction Bank (601939)}, \texttt{Bank of Communications (601328)}, and \texttt{China Merchants Bank (600036)}, are combined into a five-dimensional interval dataset. The interval for each stock is constructed using the daily highest and lowest prices, following the same procedure described in the previous section. The key steps are as follows:

\begin{itemize}
    \item Each stock is represented as a univariate interval using daily highest and lowest prices.
    \item The one-dimensional depth for each individual stock is calculated separately.
    \item The interval features are jointly integrated to compute the five-dimensional depth, which characterizes the joint centrality of all five stocks on each trading day.
\end{itemize}

This multi-dimensional extension enables us to analyze the collective depth structure and detect synchronized anomalies across the banking sector.

\subsubsection{Anomaly Detection}
Based on the five-dimensional depth values, the same anomaly detection criterion is applied. The trading days with the smallest depth values are identified as anomalous, as they reflect significant deviations from the typical joint behavior of the five stocks. These detected anomalies may correspond to market-wide events or systematic shocks affecting the overall banking sector.

\subsubsection{Visualization and Results Analysis}
\begin{figure}[ht]
\centering
\includegraphics[width=5cm]{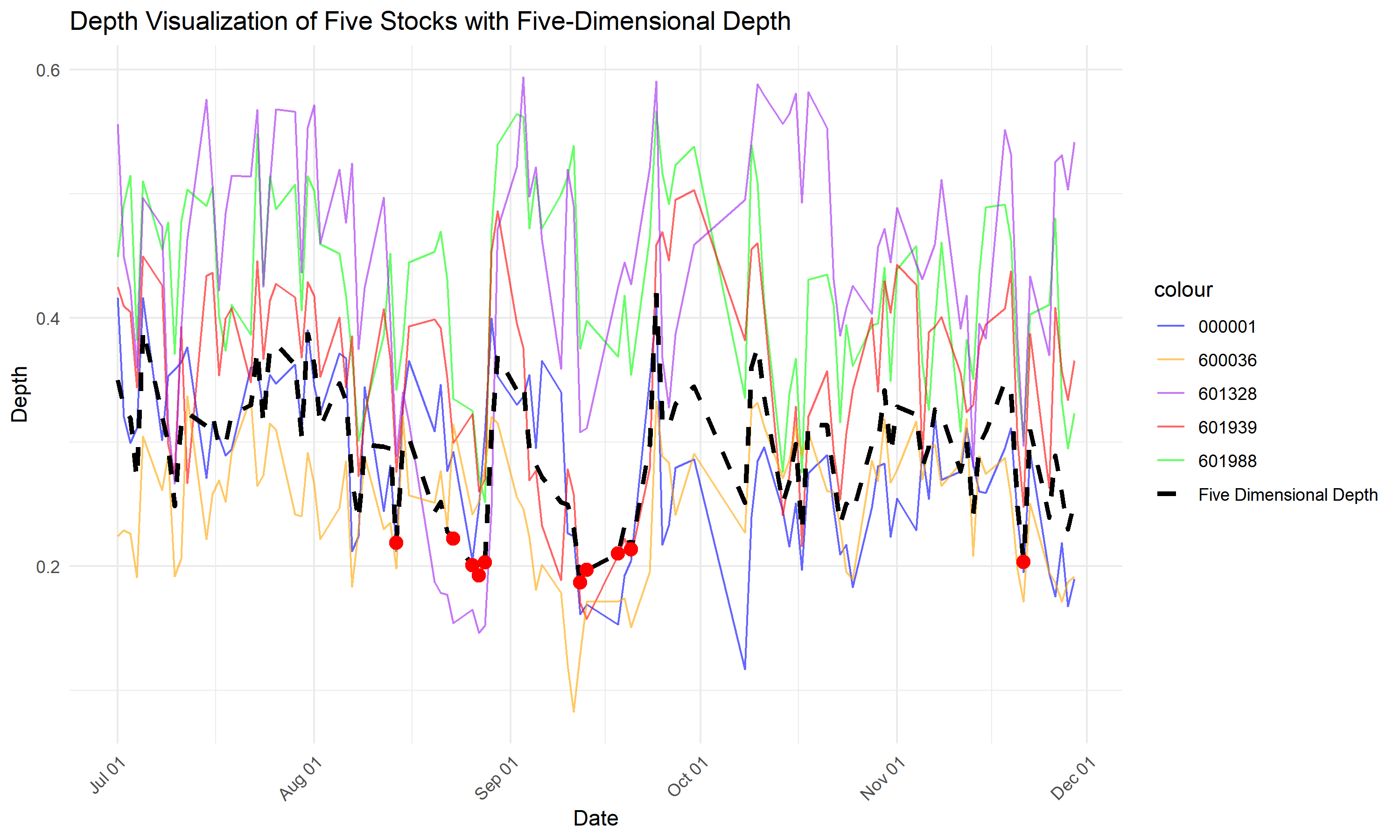}
\caption{Five-dimensional depth calculation and anomaly detection for five bank stocks. Red markers indicate the days with the smallest depth values.}
\label{fig:stock_depth_visualization}
\end{figure}

Figure \ref{fig:stock_depth_visualization} illustrates the five-dimensional depth sequence and the detected anomalies. The highlighted red markers indicate the trading days with the smallest depth values, which are recognized as anomalous days.

The results show that several dates exhibit extremely low five-dimensional depth values, corresponding to notable synchronized movements across all banks. These anomalies are likely associated with important events such as financial announcements, policy releases, or broad market shocks.

The proposed five-dimensional depth method provides a robust and holistic tool for analyzing collective market behavior. By considering multiple stocks jointly rather than individually, this approach captures cross-sectional dependencies and improves the reliability of anomaly detection in financial markets.

\subsection{GPS Trajectory Analysis}

\subsubsection{Dataset Overview}
The GeoLife GPS Trajectories dataset is a widely used spatio-temporal dataset that records the movement trajectories of 182 users over multiple years.
The dataset was collected mainly in Beijing, China, and contains 17,621 trajectory files in PLT format.
Each entry includes GPS coordinates sampled at irregular time intervals, along with timestamps, altitude, and other auxiliary information.
Such trajectory data are valuable for studying human mobility patterns and have been widely used in trajectory mining, anomaly detection, and other spatio-temporal analysis tasks.

\begin{figure}[ht]
\centering
\begin{subfigure}[t]{0.48\linewidth}
    \centering
    \includegraphics[width=\linewidth,height=3.5cm]{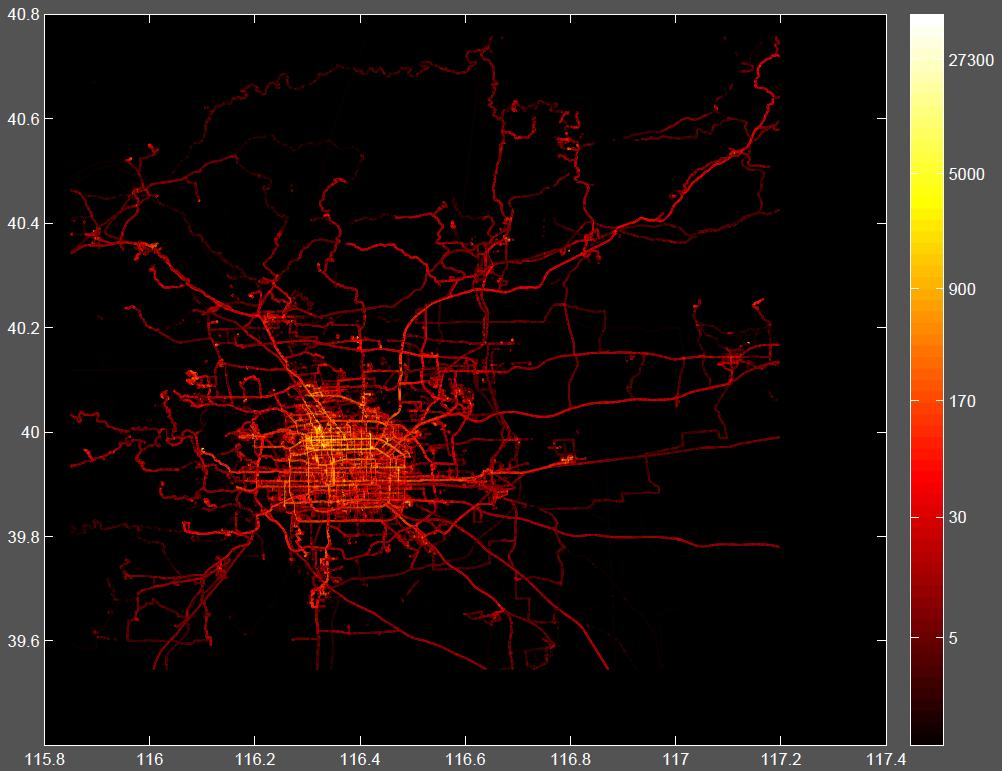}
    \caption{Data overview in Beijing}
    \label{fig:data_overview}
\end{subfigure}
\hfill
\begin{subfigure}[t]{0.48\linewidth}
    \centering
    \includegraphics[width=\linewidth,height=3.5cm]{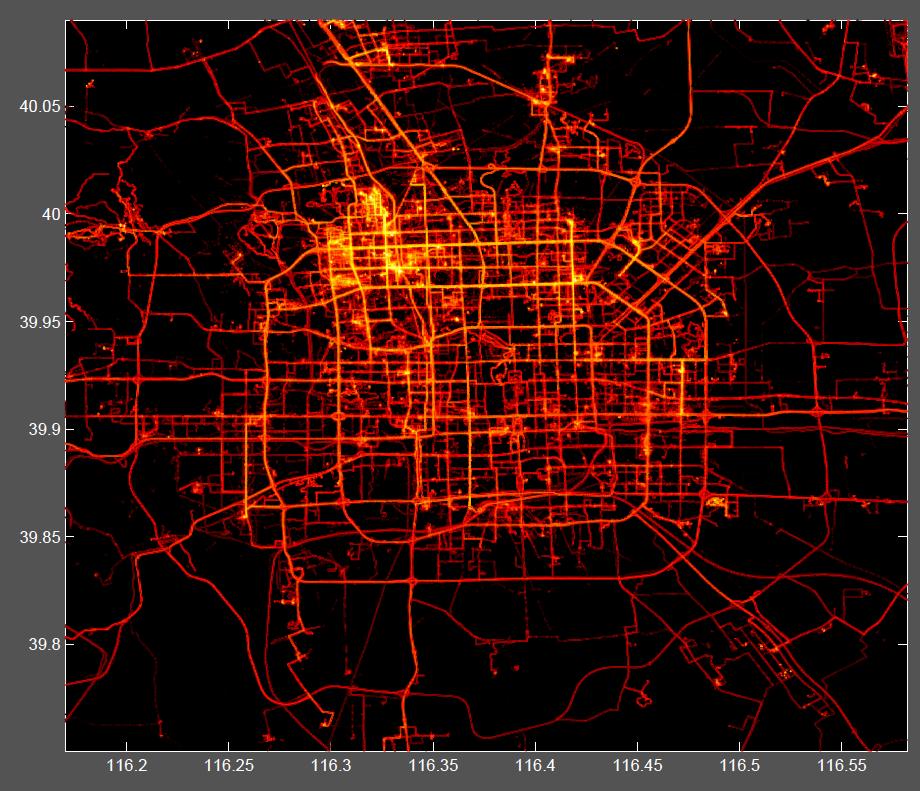}
    \caption{Within the 5th Ring Road of Beijing}
    \label{fig:within_5th_ring}
\end{subfigure}
\caption{Spatial distribution of the dataset in Beijing}
\label{fig:beijingdistribution}
\end{figure}

\autoref{fig:beijingdistribution} illustrates the spatial distribution (heatmap) of the dataset in Beijing.
The color bar indicates the density of recorded points in each region, which clearly shows the spatial coverage and concentration of GPS trajectories across the city.

\subsubsection{Data Preprocessing and Interval Representation}
In this study, we analyze trajectory data from two representative users with distinct mobility patterns.
Before constructing interval representations, strict data preprocessing is conducted:
trajectories that are excessively short, temporally discontinuous, or contain obvious positioning errors are discarded to ensure data quality.

Transforming raw GPS trajectories into two-dimensional intervals is not merely a simplification,
but also an effective way to summarize key spatial characteristics.
By representing each trajectory using its minimum and maximum latitude and longitude to form a bounding box,
we capture both the spatial extent and the general dispersion of the movement.
The latitude range $[\text{Latitude}_{\min}, \text{Latitude}_{\max}]$ and longitude range $[\text{Longitude}_{\min}, \text{Longitude}_{\max}]$
jointly define the covered region of the trajectory, offering an intuitive spatial summary.
The size of the bounding box reflects the overall spread of the trajectory: larger boxes correspond to longer or more dispersed movements.
This representation improves computational efficiency by reducing high-dimensional sequential data
to a compact and interpretable form while preserving essential spatial information.

Each valid trajectory is represented by such a two-dimensional interval,
where the latitude and longitude ranges are determined by the minimum and maximum coordinates of the trajectory.
This interval-based framework bridges raw trajectory data and depth-based statistical analysis.

\subsubsection{Anomaly Detection and Core Region Identification}
To validate the effectiveness and generalization of the proposed CLD depth method,
we conduct experiments on two typical users, namely User 91 and User 126, with diverse spatial movement habits.
Depth values are calculated independently for each trajectory interval on a per-user basis.
Trajectories with the lowest 10\% depth values are labeled as anomalies, which generally exhibit abnormal spatial coverage or offset distribution.

In contrast, the top 50\% and top 70\% intervals with higher depth scores are regarded as normal movement patterns.
We further construct their minimum bounding closures to extract stable core activity areas,
which intuitively reflect the daily routine and concentrated movement scope of each individual user.

\subsubsection{Experimental Results}
Experiments are conducted on two users with distinct mobility patterns to validate the performance of the proposed CLD method.
For each user, two groups of visual results are presented: the overall distribution of valid trajectory intervals,
and the final detection outcome containing identified anomalies as well as 50\% and 70\% core activity closures.

\begin{figure}[htbp]
\centering
\includegraphics[width=5cm]{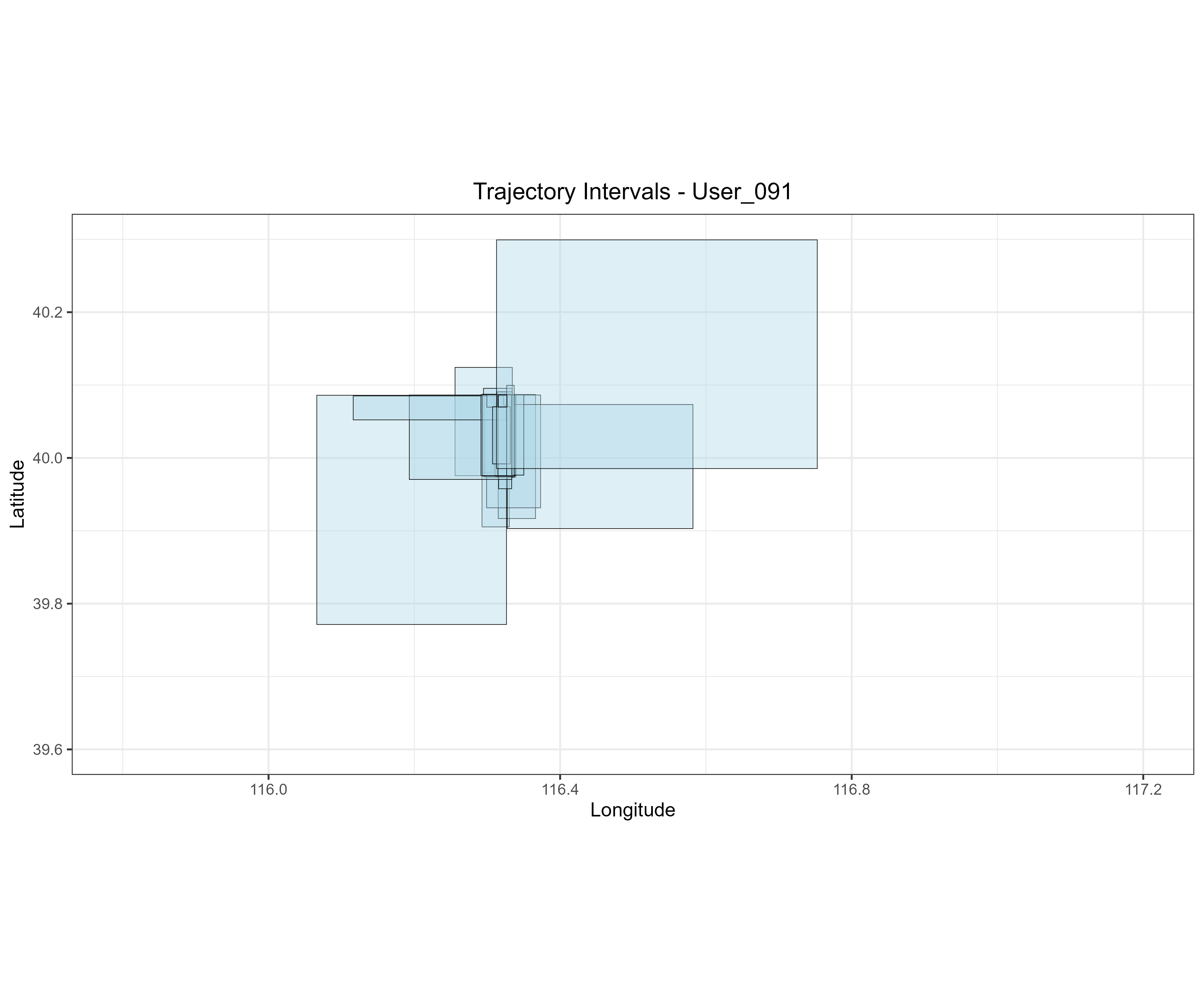}
\caption{All valid trajectory intervals of User 91 after preprocessing}
\label{fig:user91_all}
\end{figure}

For User 91, the spatial distribution of all preprocessed trajectory intervals is shown in \autoref{fig:user91_all}.

\begin{figure}[htbp]
\centering
\includegraphics[width=5cm]{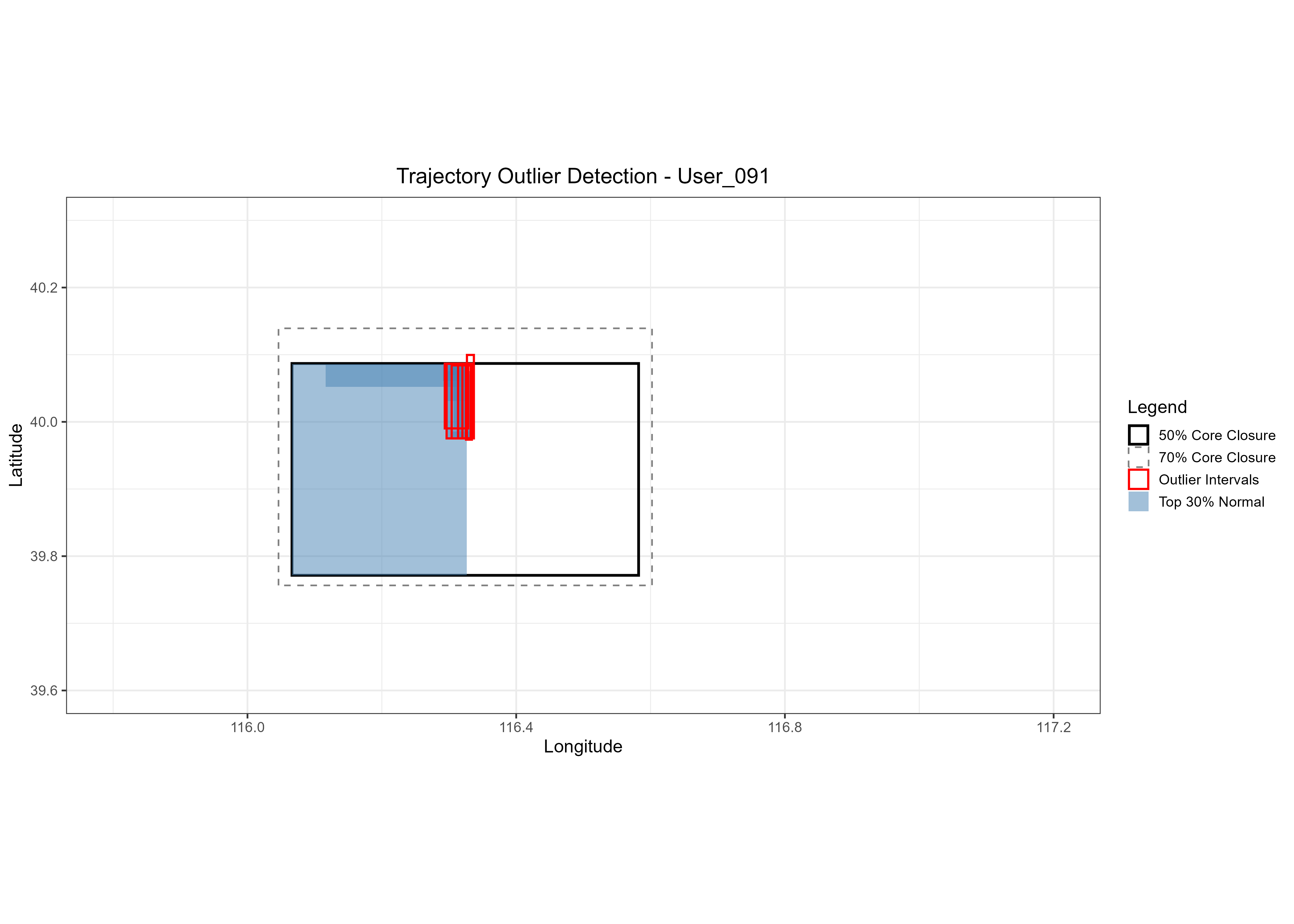}
\caption{Anomaly detection and core closures of User 91}
\label{fig:user91_final}
\end{figure}

\autoref{fig:user91_final} presents the final anomaly detection result.
The blue area marks the top 30\% normal trajectories, red boxes indicate detected anomalies,
and the black and gray dashed boxes correspond to the 50\% and 70\% core closures of frequent activity regions, respectively.

\begin{figure}[htbp]
\centering
\includegraphics[width=5cm]{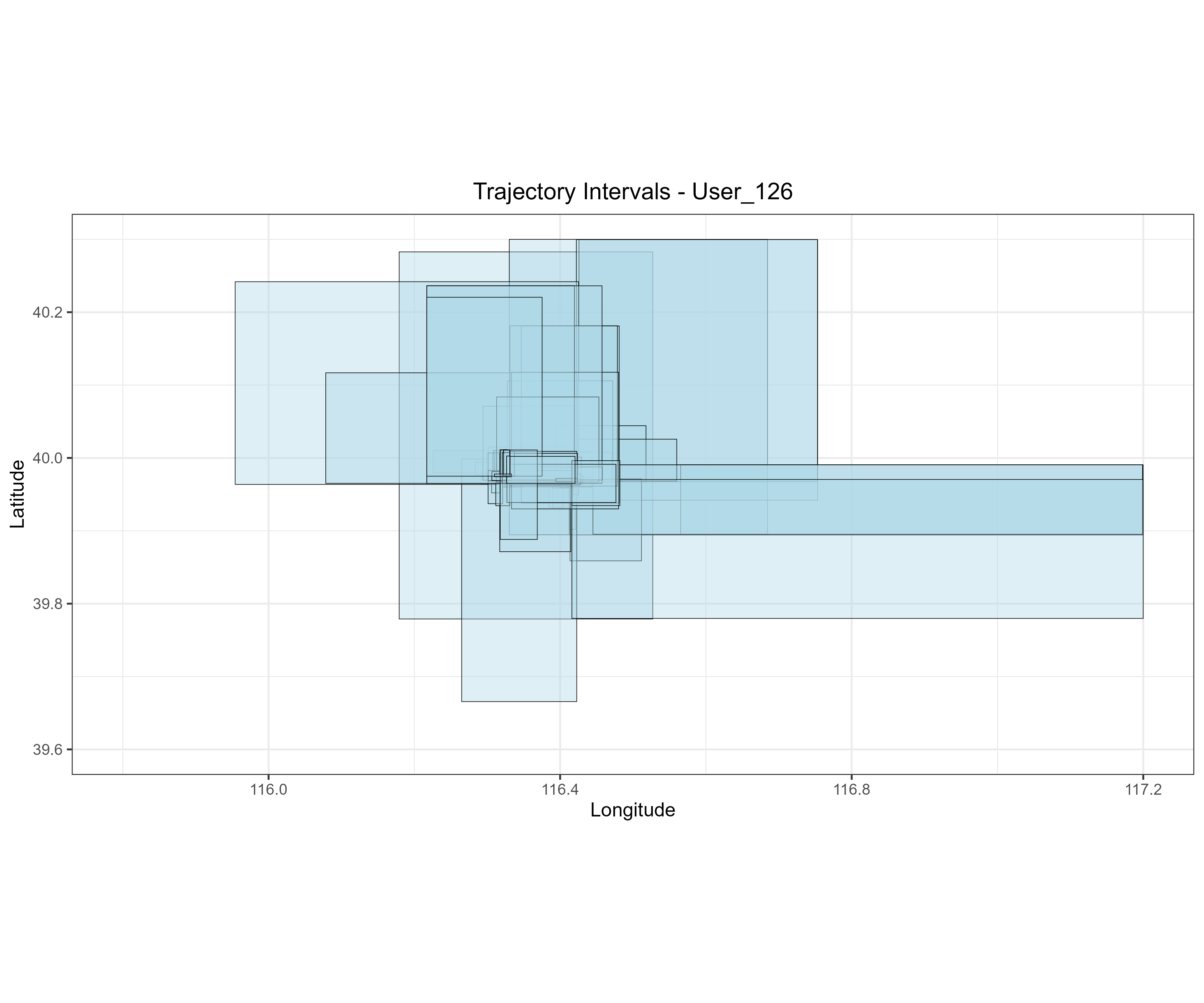}
\caption{All valid trajectory intervals of User 126 after preprocessing}
\label{fig:user126_all}
\end{figure}

For User 126, the overall distribution of processed trajectory intervals is illustrated in \autoref{fig:user126_all}.

\begin{figure}[htbp]
\centering
\includegraphics[width=5cm]{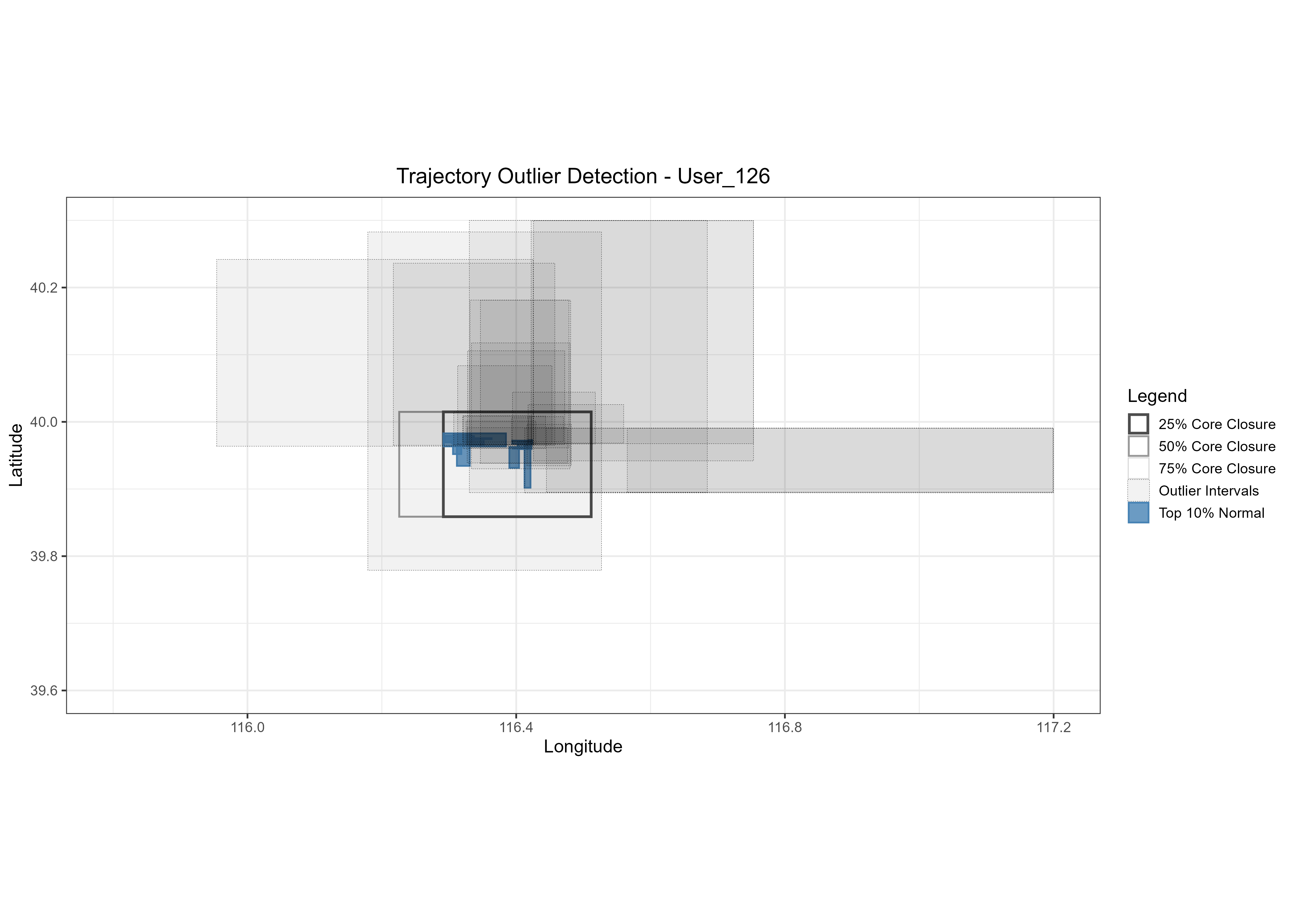}
\caption{Anomaly detection and core closures of User 126}
\label{fig:user126_final}
\end{figure}

The complete detection and core region extraction results for User 126 are displayed in \autoref{fig:user126_final}.
Even with great differences in individual movement habits, our method achieves accurate anomaly recognition and stable core activity region extraction.

Overall, the experimental results verify that the proposed CLD approach can effectively capture abnormal trajectory patterns across different users.
Core activity boundaries derived from high-depth normal intervals explicitly reflect users’ daily concentrated movement ranges.
Consistent promising results on two heterogeneous users demonstrate the good adaptability and robustness of our method in practical spatio-temporal trajectory analysis.
\section{CONCLUSION AND DISCUSSION}
In this paper, we propose a novel statistical depth for interval-valued data, using the proposed  closeness measure. The rationality and effectiveness of the proposed depth are verified through numerical simulations and real-data experiments on financial stock and GPS trajectory data. The results confirm that this method can effectively detect anomalous intervals and reveal the inherent distributional structure of interval-valued data.
This work not only enriches the theoretical framework of statistical depth for interval data, but also provides a practical tool for anomaly detection and distribution analysis in real-world applications. Nevertheless, several meaningful directions remain for future investigation. First, the method can be extended to sequential, time-varying interval data with temporal dependence to broaden its application scope. Second, it can be generalized to multivariate interval systems for joint anomaly detection and high-dimensional structural analysis. Third, computational efficiency and scalability can be improved for large-scale data via approximate computing, sampling, or parallel implementation. Finally, more domain-specific applications in medical diagnosis, environmental monitoring, and intelligent transportation deserve in-depth exploration with domain expertise.
In summary, this study provides a new perspective for interval data analysis via statistical depth. With reasonable extensions and optimizations, the proposed method can serve as a useful tool for anomaly detection, distribution analysis, and knowledge discovery from interval-valued data in diverse practical scenarios.

\appendix
\section{Proofs of CLD Core Properties}
\label{app:proofs-cld}

\subsection{Theorem A.1: Vanishing at Infinity of \(\mathrm{CLD}\)}
\begin{proof}
Let \(\mathbf{I} \in \mathcal{I}_d\) move infinitely far from \(\Omega\) in dimension \(k\). Then \(\mathrm{dist}([x_{k,1}, x_{k,2}], [x_{jk,1}, x_{jk,2}]) \to +\infty\) for all \(\mathbf{I}_j \in \Omega\), so \(\mathrm{CL}([x_{k,1}, x_{k,2}], [x_{jk,1}, x_{jk,2}]) \to 0\). Since \(\mathrm{CL}_\oplus(\mathbf{I}, \mathbf{I}_j) = \left( \prod_{k=1}^d \mathrm{CL}(\cdot, \cdot) \right)^{1/d}\) and each \(\mathrm{CL}(\cdot, \cdot) \in (0,1]\), the product tends to 0, making \(\mathrm{CL}_\oplus(\mathbf{I}, \mathbf{I}_j) \to 0\). The sample average \(\mathrm{CLD}(\mathbf{I}, \Omega)\) thus converges to 0.
\end{proof}

\subsection{Theorem A.2: Affine Invariance of \(\mathrm{CLD}\)}
\begin{proof}
For translation operator \(T\), \(T(\mathbf{I}) = ([x_{1,1}+t_1, x_{1,2}+t_1], \dots, [x_{d,1}+t_d, x_{d,2}+t_d])\). Since \(\mathrm{CL}(T(\cdot), T(\cdot)) = \mathrm{CL}(\cdot, \cdot)\), we have \(\mathrm{CL}_\oplus(T(\mathbf{I}), T(\mathbf{I}_j)) = \mathrm{CL}_\oplus(\mathbf{I}, \mathbf{I}_j)\), so \(\mathrm{CLD}(T(\mathbf{I}), T(\Omega)) = \mathrm{CLD}(\mathbf{I}, \Omega)\).

For uniform scaling operator \(S\) (\(\alpha > 0\)), \(S(\mathbf{I}) = ([\alpha x_{1,1}, \alpha x_{1,2}], \dots, [\alpha x_{d,1}, \alpha x_{d,2}])\). Similarly, \(\mathrm{CL}(S(\cdot), S(\cdot)) = \mathrm{CL}(\cdot, \cdot)\), so \(\mathrm{CL}_\oplus(S(\mathbf{I}), S(\mathbf{I}_j)) = \mathrm{CL}_\oplus(\mathbf{I}, \mathbf{I}_j)\), leading to \(\mathrm{CLD}(S(\mathbf{I}), S(\Omega)) = \mathrm{CLD}(\mathbf{I}, \Omega)\).
\end{proof}

\bibliographystyle{apalike}

\bibliography{ref} 

%\printbibliography[heading=bibintoc,title=References]

\end{document}